\documentclass{aastex631} 

\usepackage[utf8]{inputenc}
\usepackage{amsfonts}
\usepackage{amsmath}
\usepackage{natbib}
\usepackage{graphicx}

\begin{document}

\title{General Relativistic Implicit Monte Carlo Radiation-Hydrodynamics}

\author{Nathaniel Roth}
\affil{Lawrence Livermore National Laboratory, P.O. Box 808, Livermore, CA 94550, USA}

\author{Peter Anninos}
\affil{Lawrence Livermore National Laboratory, P.O. Box 808, Livermore, CA 94550, USA}

\author{Peter B. Robinson}
\affil{Lawrence Livermore National Laboratory, P.O. Box 808, Livermore, CA 94550, USA}

\author{J. Luc Peterson}
\affil{Lawrence Livermore National Laboratory, P.O. Box 808, Livermore, CA 94550, USA}

\author{Brooke Polak}
\affil{Universit{\"a}t Heidelberg, Zentrum f{\"u}r Astronomie, Institut f{\"u}r Theoretische Astrophysik, Albert-Ueberle-Str. 2, D-69120 Heidelberg, Germany}
\affil{American Museum of Natural History, 79th Street at Central Park West, New York, NY 10024, USA}

\author{Tymothy K. Mangan}
\affil{Stony Brook University, Stony Brook, NY 11794, USA}

\author{Kyle Beyer}
\affil{Department of Nuclear Engineering and Radiological Science, University of Michigan, Ann Arbor, Michigan 48109-2104, USA}

\begin{abstract}
We report on a new capability added to our general relativistic radiation-magnetohydrodynamics code,
\emph{Cosmos++}: an implicit Monte Carlo (IMC) treatment for radiation transport.
The method is based on a Fleck-type implicit discretization of the radiation-hydrodynamics equations,
but generalized for both Newtonian and relativistic regimes. A multiple reference frame approach
is used to geodesically transport photon packets (and solve the hydrodynamics equations) 
in the coordinate frame, while radiation-matter interactions are handled either in the fluid or electron frames
then communicated via Lorentz boosts and orthonormal tetrad bases attached to the fluid. We describe a method for constructing estimators of radiation moments using path-weighting that generalizes to arbitrary coordinate systems in flat or curved spacetime. Absorption, emission,
scattering, and relativistic Comptonization are among the matter interactions considered in this report. 
We discuss our formulations and numerical methods, and validate our models against a suite of 
radiation and coupled radiation-hydrodynamics test problems in both flat and curved spacetimes.
\end{abstract}

\section{Introduction}
\label{sec:Intro}

Advances made in the modern era of multi-messenger astronomy, together with ever faster computing platforms,
have inspired the development of increasingly sophisticated computational tools capable of modeling astrophysical
systems as complex as kilonovae, tidal disruption events, gamma ray bursts, and accretion disks.
These events are highly energetic interactions generally associated with compact objects (white dwarfs, neutron stars, black holes)
where relativistic effects become important and radiative emissions are critical for interpreting observational data
in the form of x-ray spectra and luminosity light curves.

Decades of progress has seen the field of computational physics advance from relatively simple N-body and hydrodynamics simulations
to the point where relativistic magneto-hydrodynamics (MHD) together with radiation treatments beyond the diffusion approximation have become
common place  \citep[a far from incomplete list includes][]{Farris2008,muller10,shibata11,zanotti11,sadowski13,mckinney2014,tominaga15,kuroda16,ryan20}.
A historical perspective of this progress is exemplified by our own contributions with the \emph{Cosmos++} code, which grew in
sophistication from a modest start in Newtonian hydrodynamics and flux-limited (grey) diffusion \citep{Anninos2003-1}, to eventually
incorporate general relativistic magneto-hydrodynamics on unstructured, adaptively refined grids \citep{Anninos2005-1}.
Many subsequent upgrades followed, including among others
grey radiation transport with M1 closure \citep{Fragile2014-1},
adaptive high order discontinuous Galerkin finite element MHD \citep{Anninos2017},
covariant relativistic molecular viscosity \citep{Fragile2018}, 
nuclear alpha-chain reactive networks \citep{Anninos2018}, 
and most recently multi-group radiation transport \citep{Anninos2020}. 
This latest multi-group transport capability was developed using the two-moment (M1) approximation, so although it
represents an important improvement over zero moment diffusion approximations, it does not compare
in fidelity to an honest Boltzmann treatment, especially in the treatment of scattering (Comptonization) events.

In this paper we describe further developments of \emph{Cosmos++} to include an Implicit Monte Carlo (IMC) treatment of radiation fully coupled with general relativistic MHD and spacetime curvature. Unlike moment closure
methodologies, Monte Carlo transport solves the full radiative transfer equation without artificial assumptions about closure relations for either the radiative intensity or flux.
This enhanced accuracy comes at the cost of random noise in the solution. Although noise can be reduced by increasing the number of Monte Carlo packets being tracked, 
it can be exceedingly expensive in terms of computing resources. 
Nevertheless, Monte Carlo remains a promising avenue of exploration for solving astrophysical problems where radiation plays an important 
role, especially because it is also useful for generating synthetic spectra of radiation escaping from systems of interest 
which can be compared to astronomical observations, including those of the vicinity of black holes where spacetime curvature is important \citep{Dolence2009,Schnittman2013,Zhang2019}.

Previous couplings of Monte Carlo radiative (or neutrino) transfer to hydrodynamics include \citet{Nayakshin2009,Abdikamalov2012,Haworth2012,Noebauer2012,Cleveland2015,Ryan2015,Roth2015-1,Tsang2015,Richers2017,Vandenbroucke2018,Smith2020}. Some of these are IMC methods, which modify the Monte Carlo radiative transfer coupling to hydrodynamics to allow for hydro time steps that are larger than the radiative cooling timescale of the fluid. The IMC methods adopted in this paper are closely related to those described in \citet{Roth2015-1}. A key difference however is that here we consider the fully general relativistic problem, whereas
\citet{Roth2015-1} only considered Minkowski spacetime and hydrodynamics equations that were only accurate to order $(v/c)^2$. 
In the process, we describe a relativistic generalization of the so-called ``Fleck factor'' 
\citep[in reference to the seminal paper on IMC,][]{Fleck1971-1}. Our goal of generalizing the Fleck factor was inspired by \citet{Gentile2011-1}, although our formula differs from the one described in those proceedings. We additionally point out that our methods differ from previous general relativistic MC rad-hydro treatments \citep[e.g.][]{Ryan2015} by the implicit discretization method developed here, the introduction of the stabilizing relativistic Fleck factor, and the step-length weighting procedure for Monte Carlo estimators. We also describe our treatment of Compton scattering, which remains valid in the high temperature regime where electrons can move relativistically in the comoving frame of the fluid. Another important aspect of this work is a discussion of the procedure for initializing Monte Carlo packets for a flow field traveling across an Eulerian grid so as to maintain zero flux in the comoving frame (section~\ref{sec:Initialization}).

The remainder of this paper is organized as follows: In sections~\ref{sec:equations} and \ref{sec:Methods} we write out the equations and discuss our numerical methods, emphasizing aspects relating to the coupling of IMC to general relativistic hydrodynamics. In section~\ref{sec:Tests} we present a series of tests validating our implementation of IMC and Compton scattering in \emph{Cosmos++}. Section~\ref{sec:Performance} compares compute efficiencies between IMC and multi-group M1. Finally, we conclude in section~\ref{sec:Conclusions}.

Most of the equations in this paper are written in units where $G = c = 1$, although in a few places we leave in factors of $c$ for clarity. We adopt the usual convention whereby Greek (Latin) indices refer to spacetime (spatial) coordinates and impose a $(-,+,+,+)$ metric signature.

\section{Basic Equations}
\label{sec:equations}

We consider two formulations of the general relativistic radiation-hydrodynamics equations,
both of which are supported by \emph{Cosmos++}:
One based on an internal energy (IE) approach used with
nonconservative artificial viscosity methods for capturing shocks,
and a second conservative total energy (TE) approach required for high resolution Godunov methods.
For both formulations we write the (non-radiative) stress energy tensor
in the following form:
\begin{equation}
T^{\alpha\beta} = T^{\alpha\beta}_{hydro} - {Q}^{\alpha\beta} =
         (\rho h + 2 P_b) u^\alpha u^\beta 
         + (P+P_b) g^{\alpha\beta} 
         - {Q}^{\alpha\beta} ~,
\label{eqn:tmn2}
\end{equation}
where magnetic fields and artificial plus molecular viscosity have been absorbed into
a single symmetric tensor representation
${Q}^{\alpha\beta} = Q^{\alpha\beta}_S + Q_B (g^{\alpha\beta} + u^\alpha u^\beta) + b^\alpha b^\beta$.
Here $\rho$ is the fluid mass density,
$h=1+\epsilon + P/(\rho)$ is the specific enthalpy,
$\epsilon = e/\rho$ is the specific internal energy,
$P$ is the fluid pressure,
$u^\alpha$ is the contravariant fluid velocity,
$b^\alpha$ is the magnetic field,
$P_b = g_{\alpha\beta}b^\alpha b^\beta/2$ is the magnetic pressure,
$Q_B$ is the bulk viscosity,
$Q^{\alpha\beta}_S$ is the symmetric shear tensor viscosity,
and $g_{\alpha\beta}$ is the curvature metric.
The fluid density, pressure, energy density, and specific energy are all defined in the comoving fluid frame.

Our implementation of radiation allows for both magnetic fields and molecular viscosity,
however for the purposes of this paper we henceforth drop ${Q}^{\alpha\beta}$ from
all subsequent discussions, but maintain occasional use of $Q_B$ representing artificial viscosity.

\subsection{IE Hydrodynamics}
\label{subsec:iehydro}

In this formulation the mass, energy, and momentum equations are derived from
$\nabla_\mu(\rho u^\mu) = S_D$,
$u_\nu \nabla_\mu T^{\mu\nu} = u_\nu G^\nu$, and 
$h_{\alpha\nu}\nabla_\mu T^{\mu\nu} = h_{\alpha\nu} G^\nu$, respectively, resulting in the following set of equations:
\begin{equation}
\partial_t D + \partial_i (D v^i) = \sqrt{-g} ~S_D ~,
\label{eqn:mass}
\end{equation}
\begin{equation}
\partial_t {\cal E} + \partial_i ({\cal E} v^i)
   + (P - Q_B) (\partial_t W + \partial_i (W v^i))
   =  -\sqrt{-g} ~u_\nu G^\nu ~.
\label{eqn:ie_energy}
\end{equation}
\begin{equation}
\partial_t S_j + \partial_i(S_j v^i) 
    + \partial_i (\sqrt{-g} ~(P-Q_B) ~g^i_j)
    = \sqrt{-g} ~T^{\mu\nu} \Gamma_{\mu\nu j} + \sqrt{-g} ~h_{j \nu} G^\nu ~,
\label{eqn:ie_momentum}
\end{equation}
where $D=W\rho$, ${\cal E}=W \rho\epsilon$,
$W=\sqrt{-g} u^0$ is the relativistic boost factor,
$v^i=u^i/u^0$ is the fluid transport velocity,
$h_{\alpha\nu} = g_{\alpha\nu} + u_\alpha u_\nu$ is the projection operator,
$S_D$ is an arbitrary source or sink for mass production that may come from chemical or nuclear activation,
$\Gamma_{\mu\nu j}$ are the curvature Christoffel symbols,
$G^\nu$ are energy and momentum source terms from radiation coupling (or any other cooling function or acceleration force),
and $S_j = \sqrt{-g} (\rho h - Q_B) u^0 u_j$ is the momentum which explicitly includes 
bulk (or artificial) viscosity in its definition, a choice we have made to simplify the evolution equations.

The radiation equations are derived from the 4-divergence of the radiation stress tensor, written simply
as $\nabla_\beta R^\beta_{\ \alpha} = -G_{\alpha} = -\int d\nu G_{\alpha(\nu)}$ when integrated over frequency.
In a coordinate frame the integrated 4-force with thermal emission and Compton scattering takes the form
\begin{equation}
G^\mu= -\rho \left(\kappa_{\mathrm{F}} + \kappa_{\mathrm{S}} \right) R^{\mu \nu } u_{\nu} 
       -\rho\left[\left(  \kappa_{\mathrm{S}} + 
                        4 \kappa_{\mathrm{S}}\left(\frac{k_B T - k_B T_R}{m_e c^2}\right) +
                          \kappa_{\mathrm{F}} - 
                          \kappa_{\mathrm{A}}\right) R^{\alpha \beta} u_{\alpha} u_{\beta} + 
                          \kappa_{\mathrm{P}} a_R T^4 \right] u^{\mu}~,
\label{eqn:4force}
\end{equation}
where $a_R$ is the radiation constant, $T$ and $T_R$ are the gas and radiation temperatures,
$\kappa_{\mathrm{S}}$ is the scattering opacity,
and $\kappa_{\mathrm{F}}$,
$\kappa_{\mathrm{A}}$, and $\kappa_{\mathrm{P}}$ are the flux, absorption, and Planck
mean opacities, respectively. Local thermodynamic equilibrium (LTE) has been assumed to relate the fluid's radiation emissivity to its absorptive opacity to radiation.

More will be said about the transport equations and coupling expressions in a later section, but it is worth pointing
out that the energy and momentum source terms in equations (\ref{eqn:ie_energy}) and
(\ref{eqn:ie_momentum}) reduce to the familiar Newtonian forms
\begin{eqnarray}
u_\nu G^\nu &=& - \rho \left(  \kappa_A +
                                    4 \kappa_{\mathrm{S}}\left(\frac{k_B T_R - k_B T}{m_e c^2}\right) \right)
                              R^{\alpha\beta} u_\alpha u_\beta
                             -\rho \kappa_P a_R T^4   ~,  \\
h_{j\nu} G^\nu &=& -\rho ( \kappa_F + \kappa_S ) R^{\alpha\beta} u_\alpha h_{\beta j}  ~,
\end{eqnarray}
when associating $R^{\alpha\beta} u_\alpha u_\beta$ with the radiation energy density, and
$R^{\alpha\beta} u_\alpha h_{\beta j}$ with the flux.

\subsection{TE Hydrodynamics}
\label{subsec:tehydro}

The conservative scheme derives the hydrodynamics equations
directly from the four-divergence of the mixed index stress tensor
$\nabla_\mu T^\mu_\nu = G_\nu$
\begin{equation}
\partial_t(\sqrt{-g} ~T^0_\nu) + \partial_i(\sqrt{-g} ~T^i_\nu)
    = \sqrt{-g} ~T^\mu_\sigma \ \Gamma^\sigma_{\mu\nu} + \sqrt{-g} ~G_\nu ~,
\end{equation}
where as before $\Gamma^\sigma_{\mu\nu}$ are the Christoffel symbols, and $G_\nu$ is the radiation 4-force coupling.
Defining total energy as ${\cal E_T} = - \sqrt{-g} {T}^0_0$ and momentum
${\cal S}_j = \sqrt{-g} {T}^0_j$, the corresponding conservation equations become
\begin{equation}
\partial_t {\cal E_T} + \partial_i({\cal E_T} ~v^i)
    + \partial_i ( \sqrt{-g} ~P ~v^i)
    = - \sqrt{-g} ~T^\mu_\sigma \ \Gamma^\sigma_{\mu 0} - \sqrt{-g} ~G_0 ~,
\end{equation}
\begin{equation}
\partial_t {\cal S}_j + \partial_i({\cal S}_j ~v^i)
    + \partial_i ( \sqrt{-g} ~P ~g^0_j ~v^i)
    = \sqrt{-g} ~T^\mu_\sigma \ \Gamma^\sigma_{\mu j} + \sqrt{-g} ~G_j ~.
\end{equation}
Mass conservation (\ref{eqn:mass}) and the 4-force expression (\ref{eqn:4force}) complete this system of equations.

\subsection{Primitive Inversion}
\label{subsec:primitive}

In addition to the conserved (evolved) fields, one must also define the
thermodynamic or equation of state variables (pressure, temperature, sound speed),
and compute so-called primitive fields (mass density, internal energy, velocity)
used in calculating the thermodynamic state, the radiation 4-force, and
the stress tensor fueling curvature flow. \emph{Cosmos++} offers numerous options
for equations of state ranging from simple ideal gas to complex (tabular) degenerate electron and nucleon models.
For Newtonian systems the extraction of primitive fields from conserved quantities
is straightforward enough, but the nonlinear coupling characteristic of relativistic hydrodynamics requires
special iterative treatment. We have implemented several fully implicit Newton-Raphson schemes for
solving the inversion problem, including reduced one, two, and five dimensional schemes for MHD, 
and a 5+4$N$ dimensional method for radiation-MHD with $N$ frequency groups.
All of these options have been described in our earlier publications \citep{Fragile2012-1,Fragile2014-1,Anninos2017,Anninos2020}, 
so we do not go into detail here, other than to summarize the list of conserved fields
(computed directly through the evolution equations),
$U^i \equiv [D, \ {\cal E_T}, \ {\cal S}_j, \ {\cal R}_0, \ {\cal R}_j] = \sqrt{-g} [u^0\rho, \ -T^0_0, \ T^0_i, \ R^0_0, \ R^0_j]$,
and the primitive fields (extracted from evolved fields via Newton iteration)
$P^j \equiv [\rho, \ \epsilon, \ \widetilde{u}^k, \ E_R, \ {u}^i_R]$.
Here $\widetilde{u}^k = u^k - u^0 g^{0k}/g^{00}$ is the velocity projected to normal
observers, and ${\cal R}_\alpha$, $R^0_\alpha$, $E_R$, ${u}^i_R$ are stand-ins for the radiation fields required
for our M1 treatment but are not necessary for our IMC implementation, an added benefit considering the computational
cost this imposes on multi-group calculations.
The cost trade-off between IMC and M1 is discussed further in section~\ref{sec:Performance}.

\subsection{Radiation Stress Tensor}
\label{subsec:radstresstensor}

The most common approaches for treating relativistic transport have
been based on one or two moment formalisms defined by a radiation stress tensor
\begin{eqnarray}
R^{\alpha \beta} &=& E n^\alpha n^\beta + F^\alpha n^\beta + F^\beta n^\alpha + P^{\alpha \beta} ~, \label{eqn:Rab_inert} \\
                         &=& \widehat{E} u^\alpha u^\beta + \widehat{F}^\alpha u^\beta + \widehat{F}^\beta u^\alpha + \widehat{P}^{\alpha \beta} ~, \label{eqn:Rab_fluid} \\
                         &=& \frac{4}{3} E_R u^{\alpha}_R u^{\beta}_R + \frac{1}{3} E_R g^{\alpha \beta} ~, \label{eqn:Rab_rad}
\label{eqn:radstress_all}
\end{eqnarray}
evaluated in inertial (\ref{eqn:Rab_inert}), comoving (\ref{eqn:Rab_fluid}), or radiation (\ref{eqn:Rab_rad}) frames. Here
$n^\alpha = \alpha^{-1}[1, ~-\beta^i]$ ($n_\alpha = [-\alpha,\ 0,\ 0,\ 0]$) is the timelike vector orthogonal to the spacelike
hypersurface, and $\beta^i$ and $\alpha = 1/\sqrt{-g^{00}}$ are the shift vector and
lapse function in the 3+1 decomposition of spacetime. The quantities $E$, $\widehat{E}$, and $E_R$ represent zero moment radiation energy 
densities in different frames, $F^\alpha$ and $\widehat{F}^\alpha$ are first moment momentum densities, 
and $P^{\alpha \beta}$ and $\widehat{P}^{\alpha \beta}$ are the anisotropic stress components
generally determined by closure relations.

Monte Carlo schemes instead evaluate the full radiation tensor (high order moments included) 
by directly sampling the invariant photon distribution function \citep{Dolence2009,Ryan2015}
\begin{equation}
f_R = \frac{dN}{d^3x d^3k} \approx \sum_n w_n \delta^3 (x^i - x^i_n) ~\delta^3 (k_j - k_{j,n})  ~,
\end{equation}
such that
\begin{equation}
R^{\alpha\beta} = \int \frac{d^3k}{\sqrt{-g} ~k^0} k^\alpha k^\beta f_R 
           \approx \sum_n \frac{w_n ~k^\alpha_n ~k^\beta_n}{\sqrt{-g} ~k^0_n} \delta^3 (x^i - x^i_n) 
           \approx \frac{1}{\sqrt{-g} \Delta^3 x} \sum_n \frac{w_n ~k^\alpha_n ~k^\beta_n}{k^0_n} ~,
\label{eqn:radstresstensor}
\end{equation}
for packets with 4-momentum $k^\alpha$ and weights
$w_k$ representing the number of photons in each packet.
The form of the last expression in equation (\ref{eqn:radstresstensor})
restricts the summation be performed over a single grid cell with coordinate volume $\Delta^3 x$. In section~\ref{sec:MomentEstimators}, we will offer an improved estimator of the components of $R^{\alpha \beta}$ based on step-length weighting of packets.

\subsection{Orthonormal Bases}
\label{subsec:tetrads}

Two frames of reference are utilized in solving the coupled radiation-hydrodynamics equations:
the coordinate and comoving (fluid) frames, the latter defining a locally Minkowski metric
$\eta_{(\alpha) (\beta)}$ attached to the timeline tangent of the fluid 4-velocity.
(We adopt the standard convention of using Greek letters to represent coordinate frames,
and parenthetically enscripted Greek letters to represent tetrad bases where indices
are raised and lowered with the Minkowski metric.)
The hydrodynamics equations and photon trajectories are evolved in the coordinate frame. However, radiation-matter
interactions are more easily evaluated in the comoving frame. Transformations between the
two are performed by defining an orthonormal tetrad attached to the fluid and using a
Gramm-Schmidt orthogonalization method to complete the tetrad. In particular we define
the first 4-vector as $e^\mu_{(0)} = u^\mu$. The remaining
three vectors are selected based on either coordinate or physical (e.g, curvature, magnetic or flow field)
alignments. 
We point out that aligning the first tetrad element to the fluid velocity is the equivalent, up to a
spatial rotation, of a Lorentz boost in special relativity.
Transformations from coordinate to tetrad and tetrad to coordinate bases are
carried out by matrix products $k^{(\beta)} = e^{(\beta)}_\alpha k^\alpha$  and the inverse
$k^{\alpha} = e^{\alpha}_{(\beta)} k^{(\beta)}$. Higher rank tensors are similarly
transformed $g^{\mu \nu} = e_{(\alpha)}^\mu e_{(\beta)}^\nu \eta^{(\alpha) (\beta)}$, using the
metric tensor as an example.

In the tetrad basis, photon 4-momenta are specified by
the frequency and spatial direction, which in spherical coordinates takes the form
\begin{equation}
k^{(\alpha)} = ~h\nu[1, ~\sin\theta\cos\phi, ~\sin\theta\sin\phi, ~\cos\theta] ~,
\end{equation}
where $h$ is Planck's constant, and $\nu$ is the frequency in the comoving fluid frame, $h\nu = -k^\alpha u_\alpha$.
When transformed to the coordinate frame ($k^{\alpha} = e^{\alpha}_{(\beta)} k^{(\beta)}$) it
takes on a more generic form $k^\alpha = h\nu[u^\alpha + \ell^\alpha]$ whose components satisfy the orthonormality
condition $u^\alpha \ell_\alpha = 0$.

The convenience of this dual frame approach is thus evident in the specification of photon 4-momenta when packets are created, which is generally performed in the fluid frame where distribution samplings are well-defined and optimally computed. But perhaps the advantage is best exemplified by considering the 4-force density derived directly from the radiation stress tensor
\begin{equation}
G^\alpha = -\nabla_\beta R^{\alpha\beta}
         = -\nabla_\beta  \int \frac{d^3k}{\sqrt{-g} ~k^0} k^\alpha k^\beta f_R 
         \propto \int \frac{d^3k}{\sqrt{-g} ~k^0} k^\alpha  
                      \left[(\nu \chi_\nu) \frac{I_\nu}{\nu^3} - \frac{\eta_\nu}{\nu^2}\right]  ~,
\label{eqn:4force_imc}
\end{equation}
where $\nu \chi_\nu$, $(\eta_\nu/\nu^2)$, and $I_\nu/\nu^3$ are the invariant extinction coefficient,
emission coefficient, and specific intensity respectively. In the fluid frame this simplifies to a simple integral over frequency and solid angle
\begin{equation}
  \widehat{G}^{(\alpha)} = \int~\int d\widehat{\nu} ~d\widehat{\Omega} ~(\chi_{\widehat{\nu}} \widehat{I}_{\widehat{\nu}} - \eta_{\widehat{\nu}}) ~\widehat{n}^{(\alpha)}  ~,
  \label{eq:ComovingG}
\end{equation}
where $\widehat{n}^{(\alpha)}$ = $(1, \widehat{\ell}^{(i)})$. These contributions to the comoving four-vector can then be transformed to the coordinate frame with $G^{\alpha} = e^{\alpha}_{(\beta)} \widehat{G}^{(\beta)}$.

Yet another advantage to this fluid frame approach for constructing $G^{\alpha}$ is that it allows a straightforward generalization of the semi-implicit treatment of radiation-fluid energy exchange via the Fleck factor \citep{Fleck1971-1} to relativistic fluids (see section \ref{sec:RelativisticFleckFactor}).
We elaborate on the calculation of $\widehat{G}^{(\alpha)}$ in sections~\ref{sec:RadHydroSourceTerms}, \ref{sec:MomentEstimators}, and \ref{sec:ComptonMethod}.

\subsection{Geodesic Transport}
\label{subsec:geodesic}

Between interaction events photon packets are propagated along geodesic trajectories by solving
the covariant transport equations in the following computationally convenient form
\begin{equation}
\frac{d x^i}{dt} = \frac{g^{i \alpha} k_\alpha}{k^0} ~;  \qquad
\frac{d k_i}{dt} = \frac{1}{2} \frac{k^\alpha k^\beta}{k^0} ~\partial_i g_{\alpha\beta} ~, \label {eqn:geod1} 
\end{equation}
for the spatial coordinates $x^i$ and momenta $k^\alpha$ of each packet,
together with the constraint $k^\alpha k_\alpha = -m^2 c^4$ (zero for null geodesics) from
which $k^0$ is derived. These equations are generic and thus applicable to static or dynamical spacetimes,
to Newtonian or general relativistic flows, and to flat space foliated with curvilinear coordinates.
When the spacetime metric is known analytically the source terms are evaluated by direct calculation (we maintain
analytic expressions for numerous metric forms and their derivatives). For dynamical spacetimes
(e.g., when solving the Einstein equations or evolving pseudopotentials) the acceleration terms
are projected from the mesh to each particle position with first or second order interpolation options.

Considering the significant cost of solving these equations numerically we have additionally incorporated a
reduced option for problems where linearized gravity is a reasonable approximation. In this special case
equations (\ref{eqn:geod1}) reduce to a single source gradient
\begin{equation}
\frac{d x^i}{dt} = \frac{k^i}{k^0} ~; \qquad
\frac{d k_i}{dt} = - \left(k^0 + \frac{k^j k_j}{k^0}\right) ~\partial_i \Phi ~, \label {eqn:geod2}
\end{equation}
where $\Phi \ll 1$ is the gravitational potential.

Some further optimization could be done even for problems involving strong gravity .
To this end we have implemented a number of different solver
options including 1st, 2nd and 4th order Runge-Kutta methods, and a velocity Verlet algorithm
which minimizes the number of metric gradient calculations.
Also, it is reasonable to expect when intervals
between interaction events (scattering, absorption, census) are much shorter than local curvature 
gradients that one can approximate photon trajectories by tangential paths calculated from the
last curvature update. For example, a measure of distance to local curvature `scatter', such as
$\min_{ijk}[g_{ij}/|\partial_k g_{ij}|]$, might be used to trigger (or deactivate) a full geodesic solve
when measured against other interaction scales.

\section{Numerical Methods}
\label{sec:Methods}

The full set of radiation-MHD equations are solved by operator splitting source terms into
spacetime advection, curvature flow (Eulerian treatment for the field equations, geodesic transport for
photon packets), constrained transport (or staggered vector potential) with magnetic fields,
radiation-matter coupling, and finally primitive field inversion. Conserved fields are
updated in time by the (user-specified) order that physics packages are inserted into a driver program. Primitive fields, from which
interaction terms are derived, are updated simultaneously at the end of each time cycle from the advanced conserved
fields after all physics packages have been evolved. 
Our numerical methods for the non-radiation parts have been described in previous publications
\citep{Anninos2005-1,Fragile2014-1,Anninos2017,Anninos2020} so we do not go into great detail here, except to
note that we use high resolution shock capturing techniques for the hyperbolic elements (including HLL and LF Riemann
solvers, together with PPM flux reconstruction), and explicit evaluation of curvature source terms.
All field equations are solved on unstructured (adaptively refined) meshes using second order finite volume
methods and up to fifth order time discretization. Primitive inversion is
solved with a fully implicit Newton-Raphson procedure after physics updates,
including radiation.

\subsection{Propagating photon packets}
\label{sec:Propagation}

Photon packets are created in the fluid frame with source-specific distribution sampling in angle and
frequency. They are transported in the coordinate frame by advancing according to the geodesic equations between scattering events. When scattering events are triggered, packets are transformed to either the fluid frame, or to electron rest frame via an additional Lorentz boost, to perform the scatter. They are then transformed back to the coordinate frame to continue along geodesic trajectories. This process repeats until the packets are destroyed by absorption or exit the grid.

We developed two strategies for updating the energy of packets due to absorption. The first decrements packet weights after each step interval in accordance with the rate of energy absorption experienced along that path. Following this strategy, packets continue to be transported until their weight falls below a designated threshold.
The second strategy avoids changing packet weights during each step. Instead, according to this strategy, we randomly compute a probability for removing the packet entirely during each interaction in proportion to the ratio of absorption opacity to total opacity, which includes the scattering contribution. This second option is consistent with interpreting a packet as a `super-photon', behaving as we would expect individual photons to behave, but transporting energy far greater than would be transported by individual photons. In addition to aiding intuition, this super-photon approach has the tendency to keep the energy of all active packets nearly equal in many problems. In contrast, the first approach (which involves decrementing packet weights) leads over time to a wide range of packet weights, as packets that were created earlier in the calculation will have their weights substantially reduced when compared to newly created packets. Generally, the efficiency of a Monte Carlo method improves as packet weights become increasingly similar \citep{Kalos2008}. For these reasons, we use the super-photon approach for all results described in this paper, but maintain the first option for comparison purposes.

According to the super-photon strategy, each time a packet is moved, we sample an interaction distance from an exponential distribution that accounts for both absorption and scattering processes, with average lengths equal to the frequency-dependent mean free path associated with the combined comoving frame absorption and scattering processes. We then transform the resulting interaction distance to a coordinate-frame distance and compare the coordinate time required for a packet to traverse that distance to the coordinate time it would take for each photon to reach a zone boundary, and to the coordinate time remaining until the end of the time step. The shortest of these timescales is used to move the photon to its next location, repeating the process as necessary until the end of the cycle.

In order to simplify the notation in the following more formal discussions we forego strict adherence to some of the index 
conventions established in the earlier theoretical sections. In particular, because much of the following
centers around the fluid frame,  we drop the parenthetical labeling of tetrad bases and, except in a few places, use hatted variables (not indices) to reference the fluid frame.

\subsection{Radiation-matter coupling}
\label{sec:RadHydroSourceTerms}

As described in section~\ref{subsec:tetrads}, the exchange of energy and momentum between radiation and matter is completely described by the four-vector $G^\mu$, whose components are most straightforwardly evaluated in the comoving frame, $\widehat{G}^\mu$. We will now describe how we can simplify the equations for $\widehat{G}^\mu$ under a set of widely applicable physical assumptions.

First, we note that thermal emission processes typically have direction vectors that average to zero in the comoving frame,
allowing these emission terms to be omitted from the radiative momentum source term if statistical conservation of momentum (rather than exact conservation) is acceptable. Second, under the assumption of local thermodynamic equilibrium, the fluid source  can be represented by the Planck function $B_{\widehat{\nu}}$. Third, when scattering processes are coherent, they do not contribute to energy exchanges in the comoving frame. Finally, similar to thermal emission, coherent scattering processes typically have outgoing direction vectors that average to zero in the comoving frame. Given these simplifications, then returning to equation~(\ref{eq:ComovingG}), we find
\begin{align}
c \widehat{G}^0 &= \int d\widehat{\nu} \int \left(\chi^a_{\widehat{\nu}} ~I_{\widehat{\nu}} - \chi^a_{\widehat{\nu}} ~B_{\widehat{\nu}} \right) d\widehat{\Omega} 
                 = \int \left(\chi^a_{\widehat{\nu}} ~c \widehat{E}_{\widehat{\nu}} - \chi^a_{\widehat{\nu}} ~4 \pi B_{\widehat{\nu}} \right) ~d\widehat{\nu}
                 = \chi^a_{E} ~c \widehat{E} - \chi^a_{p} ~c a_r T^4 ~,
\label{eq:cG0hat} \\
\widehat{G}^i   &= \int d\widehat{\nu} \int {\chi}_{\widehat{\nu}} ~I_{\widehat{\nu}} ~\widehat{n}^i ~d\widehat{\Omega}
                 = \int {\chi}_{\widehat{\nu}} ~\widehat{F}^i_{\widehat{\nu}} ~d\widehat{\nu} = \chi^i_{F} \widehat{F}^i ~.
\label{eq:Gihat}
\end{align}
In the above, we use $\chi^a_{\widehat{\nu}}$ to refer to extinction from absorption, while the undecorated $\chi_{\widehat{\nu}}$ refers to a total extinction that includes both absorption and coherent scattering. The extinctions $\chi^a_E$, $\chi^a_p$, and $\chi^i_F$ are the energy, Planck, and flux mean absorption coefficients, respectively, defined as
\begin{align}
  \label{eq:chiEdefinition}
  \chi^a_E &\equiv \frac{\int d\widehat{\nu} ~ \chi^a_{\widehat{\nu}} ~\widehat{E}_{\widehat{\nu}}} {\int d\widehat{\nu} ~\widehat{E}_{\widehat{\nu}}} ~, \\
  \label{eq:chiPdefinition}
  \chi^a_p &\equiv \frac{\int d\widehat{\nu} ~ \chi^a_{\widehat{\nu}} B_{\widehat{\nu}}} {\int d\widehat{\nu}~B_{\widehat{\nu}}} ~, \\
  \chi^i_F &\equiv \frac{\int d\widehat{\nu} ~ \chi_{\widehat{\nu}} \widehat{F}^i_{\widehat{\nu}}} {\int d\widehat{\nu}~\widehat{F}^i_{\widehat{\nu}}}  ~.
\end{align}
Compton scattering, which is noncoherent in the comoving frame, is handled separately, as we describe in section~\ref{sec:ComptonMethod}.

\subsection{Comoving-frame moment estimators}
\label{sec:MomentEstimators}

We require a way to compute the terms that appear on the right-hand sides of equations~(\ref{eq:cG0hat}) and (\ref{eq:Gihat}) based on our Monte Carlo treatment of the radiation field. We begin by considering how to construct Monte Carlo estimators of moments of the radiation field in the coordinate frame, where the packets are transported. In this section, we restore factors of $c$ which have been set to 1 by choice of units in other sections. One approach would be to use equation~(\ref{eqn:radstresstensor}). For example, the radiation energy density $E$ can be found by taking the $00$ component,
\begin{equation}
  E  \leftarrow \frac{1}{\sqrt{-g} \Delta^3 x } \sum_n w_n \left(c \, k^0_n\right) ~ ,
\end{equation}
where an arrow indicates that the Monte Carlo estimator converges to the indicated radiation moment as the number of packets approaches infinity. However, we would like to more smoothly handle situations where packets may enter and leave zones over the course of a time step that lasts a coordinate time interval $\Delta t$. Additionally, when packets change direction within a zone, we would like to incorporate this into our flux estimators in a way that uses all available packet propagation information, not just their final directions at the end of the time step. So, following \citet{Lucy1999-1} and \citet{Roth2015-1}, we sum over the steps that packets take in each zone, weighted by the coordinate time $dt$ associated with each packet step, and at the end we divide by the total time interval $\Delta t$:
\begin{equation}
  E  \leftarrow \frac{1}{\Delta^4 \cal{V}} \sum_n \sum_{dt} w_n \left(c \, k^0_n\right) \left(c \, dt \right)  ~,
  \label{eq:EstimatorCoordinateE}
\end{equation}
where $\Delta^4{\cal V} = \sqrt{-g} \Delta^3 x (c \, \Delta t)$. This approaches an invariant four-volume element with increasing space and time resolution, which lets us write $\Delta^4{\cal V} = \Delta^3 \widehat{x} (c \, \Delta \widehat{t})$, because $\sqrt{-\widehat{g}} = 1$ in the orthonormal comoving frame.

Now consider the estimator for $\widehat{E}$, the radiation energy density in the orthonormal comoving frame. This can be written in terms of step lengths $\widehat{\ell} = c \, d\widehat{t}$:
\begin{equation}
  \widehat{E}  \leftarrow \frac{1}{\left(\Delta^3 \widehat{x}\right) \left(c \, \Delta \widehat{t}\right)} \sum_{n} \sum_{d\widehat{t}} w_n \left(c \, k^{\widehat{0}}_{n} \right) \, (c \, d \widehat{t}) = \frac{1}{\Delta^4{\cal V}}  \sum_{n} \sum_{\widehat{\ell}} w_n \left(c \, k^{\widehat{0}}_{n}\right) \widehat{\ell} \, \, .
\end{equation}
Next we need a way to express $\widehat{\ell}$. In general, the quantity $\nu \chi_\nu$ is invariant. We also know that optical depth along a step, $\tau_\ell = \chi_\nu \ell$,  is invariant, since it is equal to the negative of the log of the fraction of photons that remain in the packet without interacting, and photon number is an invariant quantity. This allows us to write the following set of relations:
\begin{equation}
\widehat{\ell} = \frac{\chi}{\widehat{\chi}} \ell = \frac{\widehat{\nu}}{\nu} \ell = \frac{k^{\widehat{0}}}{k^0} \ell ~,
\end{equation}
which can be used to evaluate the estimator as a function of coordinate path length
\begin{align}
  \widehat{E}  &\leftarrow 
  \frac{1}{\Delta^4{\cal V}}  \sum_{n} \sum_{\ell} w_n \left( \frac{k^{\widehat{0}}_n}{k^0_n} \, c \, k^{0}_{n}\right) \left( \frac{k^{\widehat{0}}_{n}} {k^{0}_{n}}  \ell \right) \nonumber \\
  &= \frac{1}{\Delta^4{\cal V}} \sum_{n} \sum_{\ell} w_n  \left( c \, k^{0}_{n} \right)  \ell  \left( \frac{k^{\widehat{0}}_n}{k^0_n}\right)^2  ~.
  \label{eq:EstimatorComovingE}
\end{align}
We can compare equation~(\ref{eq:EstimatorCoordinateE}) with (\ref{eq:EstimatorComovingE}). The estimator for $\widehat{E}$ is constructed almost exactly the same way as for $E$, except for two differences: two factors of $k^{\widehat{0}}_n / k^0_n$ are included for each packet step contribution, and step coordinate time intervals are given by $\ell/c$, where $\ell$ is still a coordinate-frame length. For the special case of transforming between inertial frames, the two factors of $k^{\widehat{0}}_n / k^0_n$ have a simple physical interpretation: One factor accounts for the Doppler shift of the photon energy, while the other accounts for length contraction of the step interval.

The factor $k^{\widehat{0}}_n / k^0_n$ can be computed once per zone per hydro time step, in advance of the photon propagation, based on the tetrad transformation. Although we have not done it here, this estimator could in principle be made accurate to higher order in spacetime resolution by updating the factor of $k^{\widehat{0}}_n / k^0_n$ for each packet step as $k^0_n$ is updated during geodesic propagation, and by interpolating the fluid four-velocity within the zone to redefine the tetrad transformation at each intermediate step location.

Now we can also construct the estimator for ${\chi}_E \widehat{E}$ to use in equation~(\ref{eq:cG0hat}),
\begin{align}
  \chi_E \widehat{E}  \leftarrow \frac{1}{\left(\Delta^3 \widehat{x}\right) \left(c \, \Delta \widehat{t}\right)} \sum_{n} \sum_{d\widehat{t}} \widehat{\chi}^a_{\nu} w_n \left(c \, k^{\widehat{0}}_{n}\right) \, \left(c \, d \widehat{t} \right) &= \frac{1}{\Delta^4{\cal V}}  \sum_{n} \sum_{\widehat{\ell}} w_n \widehat{\chi}^a_{\nu} \left( c \, k^{\widehat{0}}_{n}\right) \widehat{\ell} \nonumber \\
  &=\frac{1}{\Delta^4{\cal V}}  \sum_{n} \sum_{\widehat{\ell}} w_n  \left(c \, k^{0}_{n}\right) \left( \frac{k^{\widehat{0}}_n}{k^0_n} \right)\tau_\ell ~.
  \label{eq:AbsorbedEnergyEstimator}
\end{align}
Similarly, an estimator for $\widehat{F}^i$ can be constructed as
\begin{align}
  \widehat{F}^{i} / c &\leftarrow \frac{1}{\left(\Delta^3 \widehat{x}\right) \left(c \, \Delta \widehat{t}\right)} \sum_{n} \sum_{d\widehat{t}} w_n \, (c \, k^{\widehat{i}}_n) \, \left(c \, d \widehat{t} \right)
  = \frac{1}{\Delta^4{\cal V}} \sum_{n} \sum_{\ell} w_n  (c \, k^{0}_{n})  \ell  \left( \frac{k^{\widehat{0}}_n}{k^0_n}\right)^2 \ell^{\widehat{i}}_n ~,
  \label{eq:EstimatorComovingF}
\end{align}
where $\ell^{\widehat{i}}$ is described in section~\ref{subsec:tetrads}. Note that the inclusion of $\ell^{\widehat{i}}$ in equation~(\ref{eq:EstimatorComovingF}) requires computing the comoving direction vector using the tetrad transformation each time the packet undergoes a scattering event. Finally, the estimator corresponding to equation~(\ref{eq:Gihat}) is
\begin{align}
  \widehat{G}^i  &\leftarrow \frac{1}{\left(\Delta^3 \widehat{x}\right) \left(c \, \Delta \widehat{t}\right)} \sum_{n} \sum_{d\widehat{t}} \widehat{\chi}_\nu w_n (c \, k^{\widehat{0}}_{n}) \, \ell^{\widehat{i}}_n \, \left( c \, d \widehat{t} \right)
  =\frac{1}{\Delta^4{\cal V}}  \sum_{n} \sum_{\ell}  w_n  (c \, k^{0}_{n}) \left( \frac{k^{\widehat{0}}_n}{k^0_n} \right)\tau_\ell \, \ell^{\widehat{i}}_n ~ .
\end{align}

\subsection{Relativistic Fleck factor}
\label{sec:RelativisticFleckFactor}

An important advantage to handling energy and momentum exchange between radiation and gas in the comoving frame of the fluid is 
that it allows straightforward generalization of the semi-implicit treatment of radiative energy exchange,
adapting the Fleck factor \citep[][hereafter FC1971]{Fleck1971-1} to the case of relativistic fluid flow. Based on the principle of relativistic invariance of inertial frames, we can write equation~(\ref{eq:cG0hat}) as
\begin{equation}
c \widehat{G}^0 = \rho \frac{d \epsilon}{d\tau} = \chi^a_{E} c \widehat{E} - \chi^a_{p} c a_r T^4   ~,
\label{eq:OperatorSplitRelativisticIEEB}
\end{equation}
where $d\tau$ is the interval of proper time as measured by the comoving fluid element. This equation reduces to the analogous equation for the material energy update in FC1971 when source terms aside from thermal radiation are ignored. By writing the material energy update due to radiation this way, we are applying a form of operator splitting by first computing the change in material energy solely due to absorption and emission of radiation, and then later adding this contribution as an energy source when \emph{Cosmos++} performs its flux-conservative update of the material energy. It is reasonable, therefore, to consider $\rho$ as a constant in this material energy equation.

Following FC1971, we introduce the quantities $u_r$ and $\beta$,
\begin{align}
 u_r &\equiv a_{r} T^4 ~, \\
 \beta &\equiv \frac{1}{\rho} \frac{\partial{u_r}}{\partial \epsilon} = \frac{4 a_r T^3}{ \rho c_V} ~,
\end{align}
where we have assumed the material is a perfect gas, such that $\epsilon = c_V T$ for a specific heat $c_V$ that is independent of temperature. The quantity $u_r$ should be considered as a representation of the material temperature $T$, not equal to the radiation energy density unless the radiation and the material have reached thermal equilibrium. We then have
\begin{align}
  \frac{d u_r}{d \tau} &= \frac{d u_r}{d \epsilon} \frac{d \epsilon}{d\tau}  = \rho \beta \frac{d \epsilon}{d\tau} ~,\\
  \implies \rho \frac{d \epsilon}{d\tau} &= \frac{1}{\beta} \frac{d u_r}{d \tau} ~.
  \label{eq:DeDtautoDudtau}
  \end{align}

We now wish to discretize equation~(\ref{eq:OperatorSplitRelativisticIEEB}) in time. Let $u^n_r$ refer to the value of $u_r$ at the beginning of time step $n$. We define a time-centered $\bar{u}_r = \alpha u^{n+1}_r + (1 - \alpha) u^n_r$ for a parameter $\alpha$ such that $0 < \alpha < 1$.  The time-discretized version of equation~(\ref{eq:OperatorSplitRelativisticIEEB}) is then
\begin{equation}
  \frac{1}{\beta} \frac{u^{n+1}_r -u^n_r}{\Delta \tau} = \chi^a_E c \widehat{E} - \chi^a_p  c \left[ \alpha u^{n+1}_r + (1 - \alpha) u^n_r\right] ~.
  \label{eq:DiscretizedIEEB}
\end{equation}
We can solve equation~(\ref{eq:DiscretizedIEEB}) for $u^{n+1}_r$ in terms of $u^n_r$, and plug this into the definition of $\bar{u_r}$ to obtain
\begin{equation}
\bar{u}_r = \frac{\alpha \beta c \Delta \tau \chi^a_p}{1 + \alpha \beta c \, \Delta \tau \, \chi^a_p} \, \frac{\chi_E \widehat{E}}{\chi^a_p} + \frac{1}{1 + \alpha \beta c \, \Delta \tau \, \chi^a_p} u^n_r  ~,
\end{equation}
where the common multiplier is associated with the Fleck factor
\begin{equation}
  \label{eq:RelativisticFleckFactor1}
  f_{\rm IMC} = \frac{1}{1 +  \alpha  \beta c \chi^a_p \, \Delta \tau  } ~.
\end{equation}
Notice that this formula for $f_{\rm IMC}$ utilizes the proper (not coordinate) time interval $\Delta \tau$.

In terms of $f_{\rm IMC}$, the discretized radiative heating and cooling equation can be written as
\begin{equation}
  \frac{1}{\beta} \frac{u^{n+1} -u^n}{\Delta \tau} = \chi^a_E c \widehat{E} - \chi^a_p  c \left[\left(1 - f_{\rm IMC}\right) \frac{\chi_E \widehat{E}}{\chi^a_p} - f_{\rm IMC} u^n_r \right] ~.
\end{equation}
Simplifying, and referring to equation~(\ref{eq:DeDtautoDudtau}),
\begin{equation}
  \rho \frac{\epsilon^{n+1} -\epsilon^n}{\Delta \tau} = f_{\rm IMC} \,  c \left( \chi_E  \widehat{E} -  \chi^a_p a_r T^4 \right ) ~.
\end{equation}
Comparing to equation~(\ref{eq:OperatorSplitRelativisticIEEB}), we see that to implement our semi-implicit thermal balance, all we must do is replace the true absorption extinction $\chi_{\widehat{\nu}}^a$ with an effectively reduced absorption extinction $f_{\rm IMC} \chi_{\widehat{\nu}}^a$. The Fleck factor is therefore included when we construct the absorption estimator given by equation~(\ref{eq:AbsorbedEnergyEstimator}), and it is also included when setting the energy of packets generated due to thermal emission. Moreover, the Fleck factor is included when determining the probability that a packet is absorbed during an interaction and therefore removed from the calculation. The interactions not included by the Fleck factor can be considered as a coherent effective scattering process, with extinction $(1 - f_{\rm IMC} )\chi_{\widehat{\nu}}^a$. Together the Fleck factor-weighted absorption extinction and the effective scattering extinctions sum to $\chi_{\widehat{\nu}}$, and they both contribute to the radiation force estimator defined by equation~(\ref{eq:Gihat}). 

All that remains is to use the relationship between proper time of the fluid element and coordinate time $c \, dt = u^0~d\tau$, where $u^0$ is calculated from the normalization constraint $u^\alpha u_\alpha = -c^2$.
Note that for flat spacetime this simplifies to the familiar relation $u^0 = c\gamma$ where $\gamma = 1/\sqrt{1- v_i v^i/c^2}$ is the Lorentz factor.
The final form of the Fleck factor can be written
\begin{equation}
f_{\rm IMC} = \frac{1}{1 + 4 \alpha c \chi^a_p \, (c \, \Delta t/u^0) ~a_r T^3 /(\rho c_V) } ~.
\label{eq:RelativisticFleckFactor2}
\end{equation}
We point out that this result differs from \citet{Gentile2011-1}, who used $\gamma \Delta t$ rather than $\Delta t / \gamma$ in the denominator of the formula for $f_{\rm IMC}$ in the case of flat spacetime. Both expressions lead to a stable energy update, although the \citet{Gentile2011-1} form effectively increases the coefficient $\alpha$ compared to ours, leading to less rapid cooling than can be achieved stably with our version. Our formula also generalizes to curved spacetimes, considering $u^0$ encodes metric information.

In practice, to ensure stability for problems where a nontrivial Fleck factor is required, the parameter $\alpha$ is set to a value between 0.5 and 1.0. Higher values of $\alpha$ lead to smoother solutions and improved stability. However, higher values of $\alpha$ can also lead to artificially long cooling times for zones undergoing rapid cooling. This is evident in the radiative equilibrium test described in \citet{Noebauer2012} and \citet{Roth2015-1}. 

\subsection{Initializing packets in relativistic flows}
\label{sec:Initialization}

At times it is desirable to begin a radiative transfer calculation when there is already a non-negligible field present. To that end we work out an initialization procedure for radiation packets being transported along with a moving fluid across a Eulerian grid representing an inertial frame. For pedagogical purposes, we restrict our attention to one-dimensional flows with constant velocity and uniform density and temperature in flat spacetime. Through appropriate coordinate transformations, the same procedure can be applied in more general settings. 

To help validate properties of our implementation described later in this section, we refer to the following exact relations between inertial frame moments. These relations can be obtained from the Lorentz covariance of the 
radiation stress-energy tensor, and in 1D takes the form \citep[equations 91.13--91.15 of][]{Mihalas1984-1}:
\begin{eqnarray}
\label{eq:Etransform}
     E &=& \gamma^2 \left[\widehat{E} + 2\beta c^{-1} \widehat{F} + \beta^2 \widehat{P}\right] ~,
     \label{eq:Ftransform} \\
     F &=& \gamma^2\left[ \left(1 + \beta^2\right) \widehat{F} + v \widehat{E} + v \widehat{P}\right] ~, 
     \label{eq:Ptransform}  \\
     P &=& \gamma^2 \left[\widehat{P} + 2\beta c^{-1} \widehat{F} + \beta^2 \widehat{E}\right] ~,
\end{eqnarray}
where we have simplified notation to drop vector and tensor indices.
Here, $v$ is the (potentially relativistic) velocity as measured in the inertial frame, 
and $\beta$ and $\gamma$ have their usual special relativistic meanings. 

For the special case of uniform flow, symmetry requires $\widehat{F} = 0$ and $\widehat{P} = \widehat{E}/3$.
Substituting these conditions into equations (\ref{eq:Etransform})--(\ref{eq:Ptransform}) we obtain
 \begin{align}
 \label{eq:ZeroFhatMomentTransformationsE}
     E &= \widehat{E} \left[ \gamma^2 \left(1 + \frac{1}{3} \beta^2 \right) \right] ~, \\
     F &= \left(v \widehat{E} \right) \left (\frac{4}{3 + \beta^2} \right) =
          \left(v \widehat{E} \right) \left ( \frac{4}{3} \gamma^2 \right) ~, \label{eq:ZeroFhatMomentTransformationsF} \\
     P &= \widehat{P}  \left[ \gamma^2 \left(1 + 3 \beta^2 \right) \right] =
          \gamma^2 \widehat{E} \left(\frac{1}{3} + \beta^2\right)  ~. \label{eq:ZeroFhatMomentTransformationsP}
 \end{align}

\subsubsection{Enforcing zero flux in the fluid frame}

The option to sample packet directions isotropically in the comoving frame is generally possible, even for relativistically
moving fluids, so long as we choose packet energies appropriately to ensure $\widehat{F} = 0$. In this section, we will consider packet energies $\epsilon_n = w_n (c \, k^0_n)$. One procedure we can follow to enforce this condition is to introduce a reference packet energy $\epsilon_{\rm ref}$, whose value we will determine shortly, and set lab-frame packet energies as 
\begin{equation}
\label{eq:IntroduceEref}
\epsilon_n = \left(\frac{\widehat{\nu}}{\nu} \right)_n^{-2} \epsilon_{\rm ref} ~.
\end{equation} 
Here we have identified $k^{\widehat{0}}_n/k_n^0$ with $(\widehat{\nu} / \nu)_n$, the ratio of frequencies measured in the comoving frame and the lab frame, respectively. Then, starting from equation~(\ref{eq:EstimatorComovingF}), and taking the limit $\Delta t \to 0$ so that ${\ell}/( c \Delta {t}) \to 1 $, we have
\begin{equation}
\widehat{F}^i/c \leftarrow \frac{1}{\Delta V} \sum_n \epsilon_n \,  \left( \frac{\widehat{\nu}}{\nu} \right)^2_n  \widehat{\ell}_n^i 
                       = \frac{1}{\Delta V} \epsilon_{\rm ref} \sum_n \widehat{\ell}_n^i \to 0 \, \, ,
\end{equation}
where the last sum approaches zero because of our choice to initialize packet directions isotropically in the comoving frame. 

We next relate $\epsilon_{\rm ref}$ to the lab-frame moment $E$ in a zone,
\begin{equation}
E \Delta V = \sum_n \epsilon_n
           = \epsilon_{\rm ref} \sum_n \left(\frac{\widehat{\nu}}{\nu} \right)_n^{-2}.
\end{equation}
To determine how this sum converges, we again exploit the fact that we have chosen the packet directions isotropically in the comoving frame. If we originally chose, say, the $x$-axis to be parallel to the fluid flow in the lab frame, then after transforming to the comoving frame we have a new axis $x^\prime$, along with the unchanged $y$ and $z$ axes. 
Let $\widehat{\mu}$ represent the cosine of the angle between the packet direction and the $x^\prime$ axis. 
Then with the  inverse Doppler relation
\begin{equation}
  \left(\frac{\widehat{\nu}}{\nu} \right)^{-1} = \gamma \left(1 + \beta ~\widehat{\mu}\right)  ~,
  \label{eq:DopplerRelation}
\end{equation}
the angle probability distribution $d{\cal P}/d\widehat{\mu} = 1/2$, and letting $N$ denote 
the number of packets in the zone at the initialization time, we can write
\begin{equation}
    \frac{1}{\epsilon_{\rm ref}} \frac{E \Delta V}{N} = 
    \frac{ \sum_n \left(\frac{\widehat{\nu}}{\nu} \right)^{-2}_n}{N} \to \int  \left(\frac{\widehat{\nu}}{\nu} \right)_p^{-2} \frac{d \cal{P}}{d\widehat{\mu}} d\widehat{\mu} 
     = \frac{1}{2} \int_{-1}^{1} \gamma^2 \left(1 + \beta ~\widehat{\mu} \right )^2 d\widehat{\mu}
     = \gamma^2 \left( 1 + \frac{1}{3} \beta^2 \right)   ~.
\end{equation}
We recognize the final expression as part of equation (\ref{eq:ZeroFhatMomentTransformationsE}), namely $E/\widehat{E}$. 
So we now have a formula for $\epsilon_{\rm ref}$ in terms of the comoving radiation moment,
\begin{equation}
\epsilon_{\rm ref} = \frac{\widehat{E} \Delta V}{N}  ~,
\end{equation}
which can be evaluated if $\widehat{E}$ is viewed as part of the initial conditions of the problem 
(if for example radiation is thermalized, $\widehat{E} = a_r T^4$).

Referring to equation (\ref{eq:IntroduceEref}), we find
\begin{equation} 
\widehat{\epsilon}_n = \left(\frac{\widehat{\nu}}{\nu} \right)_n \epsilon_n = \left(\frac{\widehat{\nu}}{\nu} \right)^{-1} \epsilon_{\rm ref}.
\end{equation}
We have arrived at what first glance appears to be a counter-intuitive result. 
Although we have given our packets an isotropic direction distribution in the comoving frame, these packets must be given \emph{unequal} energies in order to ensure zero flux in the comoving frame. Referring to the Doppler relation in equation~(\ref{eq:DopplerRelation}), we see that the energy is adjusted based on the packet direction with respect to the material motion. Or, viewed another way, we could force the comoving packets to have equal energy, but then we would have to sample their direction \emph{anisotropically}. The necessity of such a procedure was also identified in \citet{Ryan2015}, through consideration of the invariant four-volume element.

We know from symmetry that there is no preferred direction in the comoving frame. In fact, we have relied on this to derive our results. So why are packets not treated symmetrically when we initialize them in the comoving frame? The seeming contradiction is resolved when we remember that we are constructing an estimator for the comoving flux assigned to a zone whose boundaries are specified in the lab frame. In the comoving frame, these boundaries are moving at the speed of the fluid, thereby breaking the symmetry. The packets are indeed directed isotropically in the comoving frame, but the flux of packets through the zone boundaries over a proper time interval is biased in the direction anti-aligned with the apparent motion of the zone boundaries. Moreover, the initialization of packets relies on a notion of simultaneity according to the lab-frame clock. In the comoving frame, this simultaneity is broken: according to the comoving frame clock, the zone boundaries will not simultaneously reach their spacetime positions corresponding to the lab frame $t=0$ positions. The above procedure accounts for these effects and leads to proper initialization to ensure zero comoving flux in the zone. We use this procedure in the radiating shock tube tests described in section~\ref{sec:RelativisticRadiatingShocks}, and we confirm that it eliminates spurious transient behavior at early times in the regions of uniform flow when compared to a procedure that initializes packets with equal energies as measured in the comoving frame and isotropically sampled directions in the comoving frame.

\subsection{Relativistic Compton scattering}
\label{sec:ComptonMethod}

Our treatment derives from \citet{Canfield1987-1} and \citet{Dolence2009}, both of which were substantially influenced by \citet{Pozdnyakov-1983}. We briefly summarize the method here. 

For each scattering event, after having transformed into the fluid frame, we begin by sampling the speed $v_e$ of thermal electrons from
the relativistic Maxwell-Juttner distribution, given the fluid temperature.
Next we sample a direction for the electron in the form of a unit vector $\widehat{n}_e^i$. 
This direction is characterized by an angle $\theta$ between the fluid-frame photon propagation vector and the electron velocity vector, and a uniformly distributed azimuthal angle $\phi$ that is independent of $\theta$. The distribution for $\theta$ is weighted such that electrons with momenta anti-aligned with the photon momentum are more common than those that are aligned, as described by the distribution $( 1 - \beta \cos \theta)/2$, where $\beta = |v_e| / c$. For discussions of this angular dependence of the cross section, see for example \citet{Gould1971-1}.

Once we have obtained an electron velocity vector following these steps, we perform a Lorentz boost into the electron rest frame, and there we compute the angle-integrated scattering cross section $\widehat{\sigma}$ using the Klein-Nishina formula, which depends on the frequency of the photon as measured in the electron rest frame. We now apply a rejection sampling method by generating a random unit variate and comparing it to the ratio $\widehat{\sigma}/\sigma_T$ where $\sigma_T$ is the Thomson scattering cross-section $\sigma_T$ and which is the maximum possible value for $\widehat{\sigma}$. If the variate is less than this ratio we proceed, but if it is greater we reject the electron velocity vector and begin again by sampling a new $v_e$ and $\widehat{n}_e^i$. Once we accept an electron velocity, we sample the outgoing photon direction from the Klein-Nishina differential scattering cross-section, accounting for the change in photon energy due to electron recoil. We then inverse Lorentz transform the photon four-momentum back into the fluid frame, and perform one final transformation from the tetrad basis to the coordinate frame (or
inverse Lorentz boost for Minkowski spacetimes).

During each scattering event, we record the difference of the incoming and outgoing photon energy and momentum components as measured in the comoving frame, and sum these contributions over all scattering events in each zone over the hydrodynamic time step. The resulting energy and momentum sums per cell-volume per time interval are added to our baseline estimators for $\widehat{G}^\alpha$ described in sections~\ref{sec:RadHydroSourceTerms} and \ref{sec:MomentEstimators} that accounted for thermal absorption and emission, and coherent scattering.

During the photon propagation step we must also determine the fluid-frame mean free path for Compton scattering off the thermal distribution of electrons, which will be a function of the local electron temperature $T_e$ and the fluid-frame photon frequency $\widehat{\nu}$. This requires taking the average cross-section as measured in the fluid frame that results from the above sampling procedure over the distribution of electron velocity vectors. The resulting average cross-section can be expressed as a double integral over all possible values for $v_e$ and $\theta$. 
For computational expediency, we pre-compute these average cross-sections and store them 
into a two-dimensional look-up table binned logarithmically by electron temperature (ranging from $10^7$ to $10^{12}$ Kelvin) and photon energy (ranging from $h\widehat{\nu}$ = 1 eV to $10^8$ eV).

\subsection{Methods overview}
\label{sec:methodSummary}

Many steps go into a single cycle update, so we take the opportunity here to summarize a broad view of the algorithm.
First we point out that (1) none of the new material discussed in this report affect the other physics packages, except for the total energy and
momentum modified by the radiation 4-force, and (2) all of the source terms used in updating the radiation
fields are derived explicitly from current state primitive fields (only at the very end of the cycle are primitive
fields updated). Because primitive fields are not altered intermittently between physics solves, the radiation transport
solver is not affected by and proceeds (within a single cycle) essentially independently of other packages, and vice versa.

Each cycle begins with the interaction-sampling propagation sequence described in section \ref{sec:Propagation}:
Between events packets are transported by the geodesic equations; at event sites packet energies and velocities
are updated in accordance with the associated event (e.g., absorption, physical scattering, Fleck scattering).
An interpolated weighting procedure calculates comoving moment estimators (section \ref{sec:MomentEstimators}) 
and accumulates updates to the radiation 4-force,
equations (\ref{eq:cG0hat}) and (\ref{eq:Gihat}), integrating over
energies and solid angles, accounting for both absorption and scattering. At the end of each propagation cycle,
when particles reach census, the integrated 4-force is used to compute the fluid internal energy semi-implicitly with the Fleck procedure
described in section \ref{sec:RelativisticFleckFactor}.
For the total energy formulation (section \ref{subsec:tehydro}), the internal energy computed at this stage does not
replace the actual primitive field, but is instead used to update the total energy and momentum conserved fields.
After all physics packages have been advanced, primitive fields are obtained from the conserved (evolved) fields by Newton iteration before exiting the cycle.

\section{Tests}
\label{sec:Tests}

\subsection{Radiating shocks}
\label{sec:RelativisticRadiatingShocks}

This series of radiation wave and shock tube tests was first performed by \citet{Farris2008} in the Eddington closure
approximation, $\widehat{P}^{ij} = (\widehat{E}/3) ~\delta^{ij}$.
Versions of these tests have since been repeated with several other codes and treatments of the radiation, 
including our own earlier work with single-group Eddington \citep{Fragile2012-1} and multi-group M1 closure \citep{Anninos2020}.
We perform these tests again with implicit Monte Carlo, comparing results to our M1 treatment. 
Similar tests were carried out by \citet{Ryan2015} in their explicit treatment of Monte Carlo transport, which
requires smaller time steps to resolve the radiation cooling time scale and to maintain stability.

The boundary conditions for this test make use of the fact that far from the shock transition region, $\widehat{F}^i \to 0$. In the lab frame, however, absent any injection of packets from the boundaries, there is a non-zero flux through the boundaries due to the advection of radiation. In this situation, \citet{Gentile2011-1} show that the proper boundary condition requires injecting packets from the boundary with the following energy flux measured in the coordinate frame:
\begin{equation}
  F_{\rm face} = \frac{a_r c}{4} T^4\, \gamma^2 F_{\rm norm} ~ ,
  \label{eq:BoundaryFluxInjection}
\end{equation}
where
\begin{equation}
F_{\rm norm} =   \left[1 + \frac{8}{3} \left(\frac{v^i n_i}{c}\right) + 2\left(\frac{v^i n_i}{c}\right)^2 - \frac{1}{3} \left(\frac{v^i n_i}{c}\right)^4 \right ] ~ ,
  \end{equation}
and where $n^i$ is the normal vector to the boundary plane pointing into the computational domain. In this 1D problem, the boundary planes coincide with planes of constant $x$-coordinate, and we have assumed the material has no motion in the $y$ or $z$ directions, although \citet{Gentile2011-1} also discuss how to accommodate such motion. By injecting packets at a rate consistent with equation (\ref{eq:BoundaryFluxInjection}), the net radiative flux near the boundary as measured in the comoving frame will tend toward zero. Moreover, we also require the distribution of direction cosines $\mu$ with respect to the boundary normal direction for those packets injected from the boundary, measured in the coordinate frame. This will allow the packet direction distribution near the boundary to match the distribution in the interior zones. Again quoting from \citet{Gentile2011-1}, and assuming no material motion in the $y$ or $z$ directions, this distribution is
\begin{equation}
 \frac{ d {\cal P}}{d \mu} = \frac{1}{\pi \gamma^6\,  F_{\rm norm}}\frac{\mu}{\left[1 - \left(v^i n_i / c \right) \mu \right]^4}   ~.
  \label{eq:BoundaryFluxDirections}
\end{equation}

We consider five tests: a nonrelativistic strong shock (case 1), a relativistic strong shock (case 2),
a relativistic wave (case 3), a radiation pressure dominated relativistic wave (case 4), and an optically thick
variant of case 4. The parameters and initial states for all cases are summarized in Table \ref{tab:shocktube}. All calculations were performed with 128 zones covering the entire domain for both the IMC and M1 calculations, and run until transients exited the grid and the solution achieved steady-state. 
\begin{deluxetable}{ccccccccccc}
\tablecaption{Radiation Shock Tube Parameters \label{tab:shocktube}}
\tablewidth{0pt}
\tablehead{
\colhead{Case} & $\Gamma$ & $\kappa^a$ &
$\rho_L$       & $P_L$    & $u_L^x$    & $E_L$      &
$\rho_R$       & $P_R$    & $u_R^x$    & $E_R$
}
\startdata
1 & 5/3  & 0.4
  & 1    & $3   \times 10^{-5}$  & $0.0015$              & $1    \times 10^{-8}$
  & 2.4  & $1.61\times 10^{-4}$  & $6.25\times10^{-3}$   & $2.51 \times 10^{-7}$ \\
2 & 5/3  & 0.2
  & 1    & $4 \times 10^{-3}$    & 0.25                  & $2    \times 10^{-5}$
  & 3.11 & $0.04512$             & $0.0804$              & $3.46 \times 10^{-3}$ \\
3 & 2    & 0.3
  & 1    & $60$                  & 10                    & 2
  & 8    & $2.34 \times 10^{3}$  & $1.25$                & $1.14 \times 10^{3}$ \\
4 & 5/3  & 0.08
  & 1    & $6 \times 10^{-3}$    & 0.69                  & 0.18
  & 3.65 & $3.59 \times 10^{-2}$ & $0.189$               & $1.30$ \\
5 & 5/3  & 0.7
  & 1    & $6 \times 10^{-3}$    & 0.69                  & 0.18
  & 3.65 & $3.59 \times 10^{-2}$ & $0.189$               & $1.30$ 
\enddata
\end{deluxetable}

\begin{figure*}[htb!]
\begin{tabular}{cc}
\includegraphics[width=0.48\textwidth]{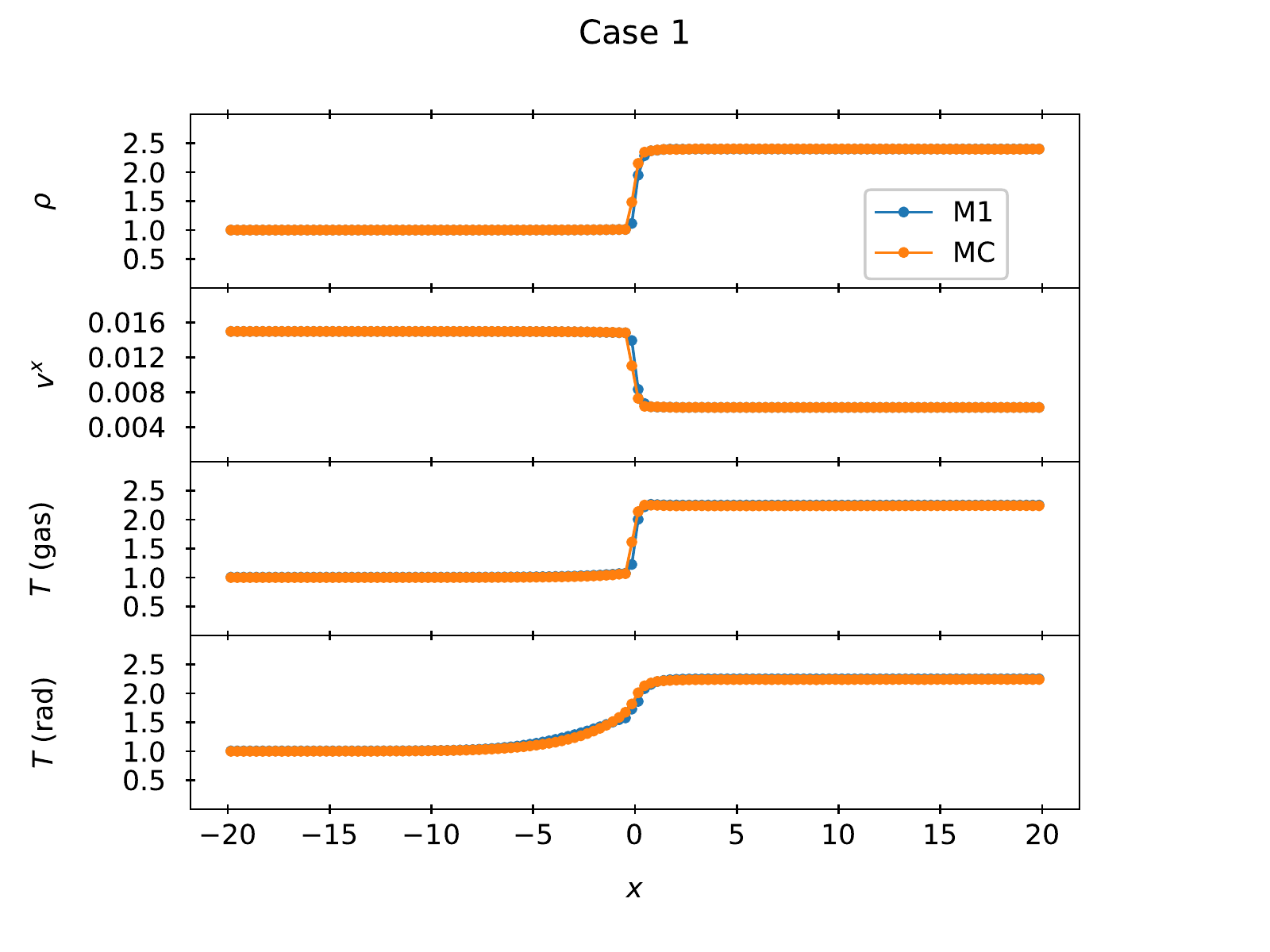} &  
\includegraphics[width=0.48\textwidth]{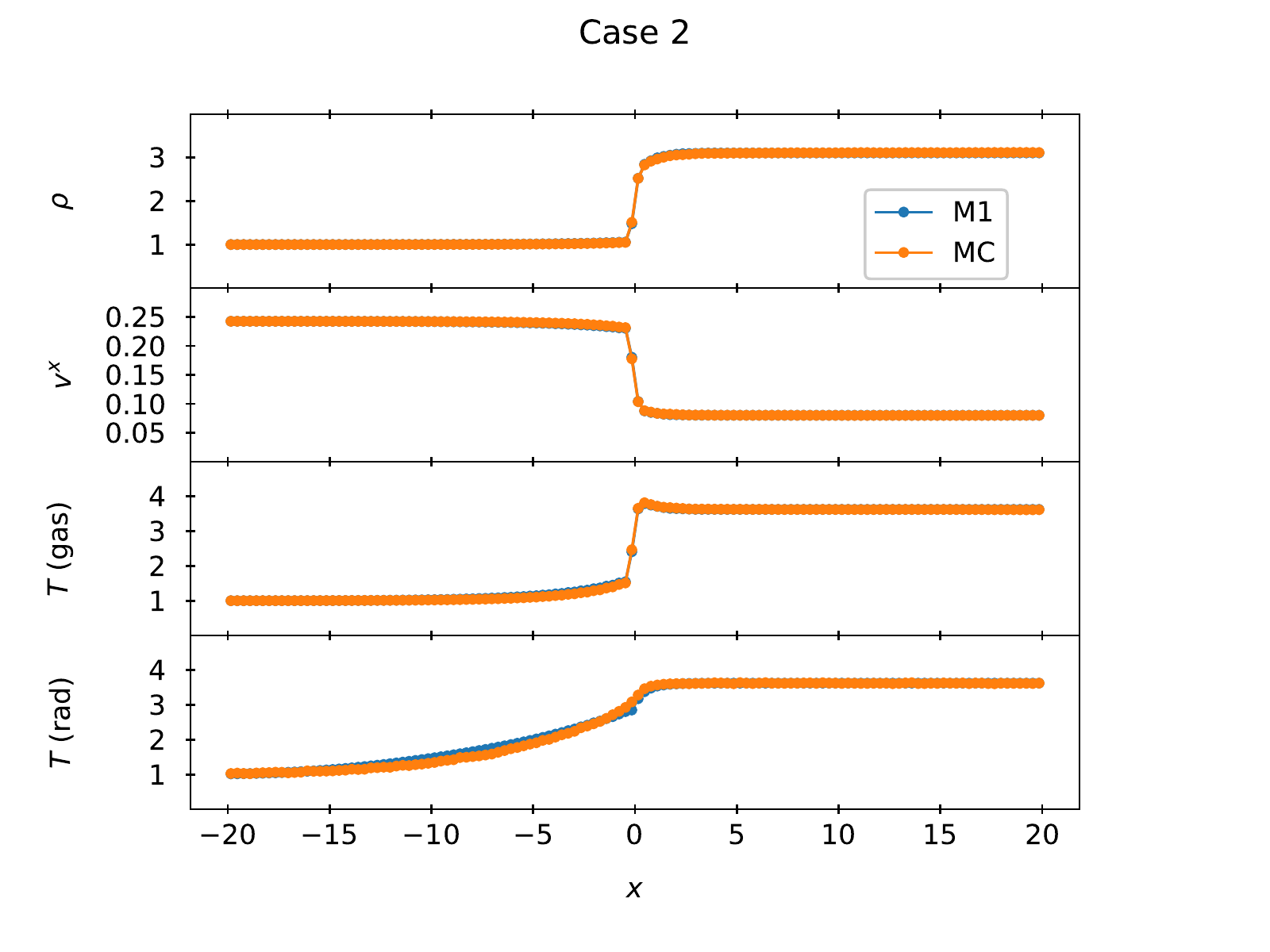} \\
(a) & (b) \\ \\
\includegraphics[width=0.48\textwidth]{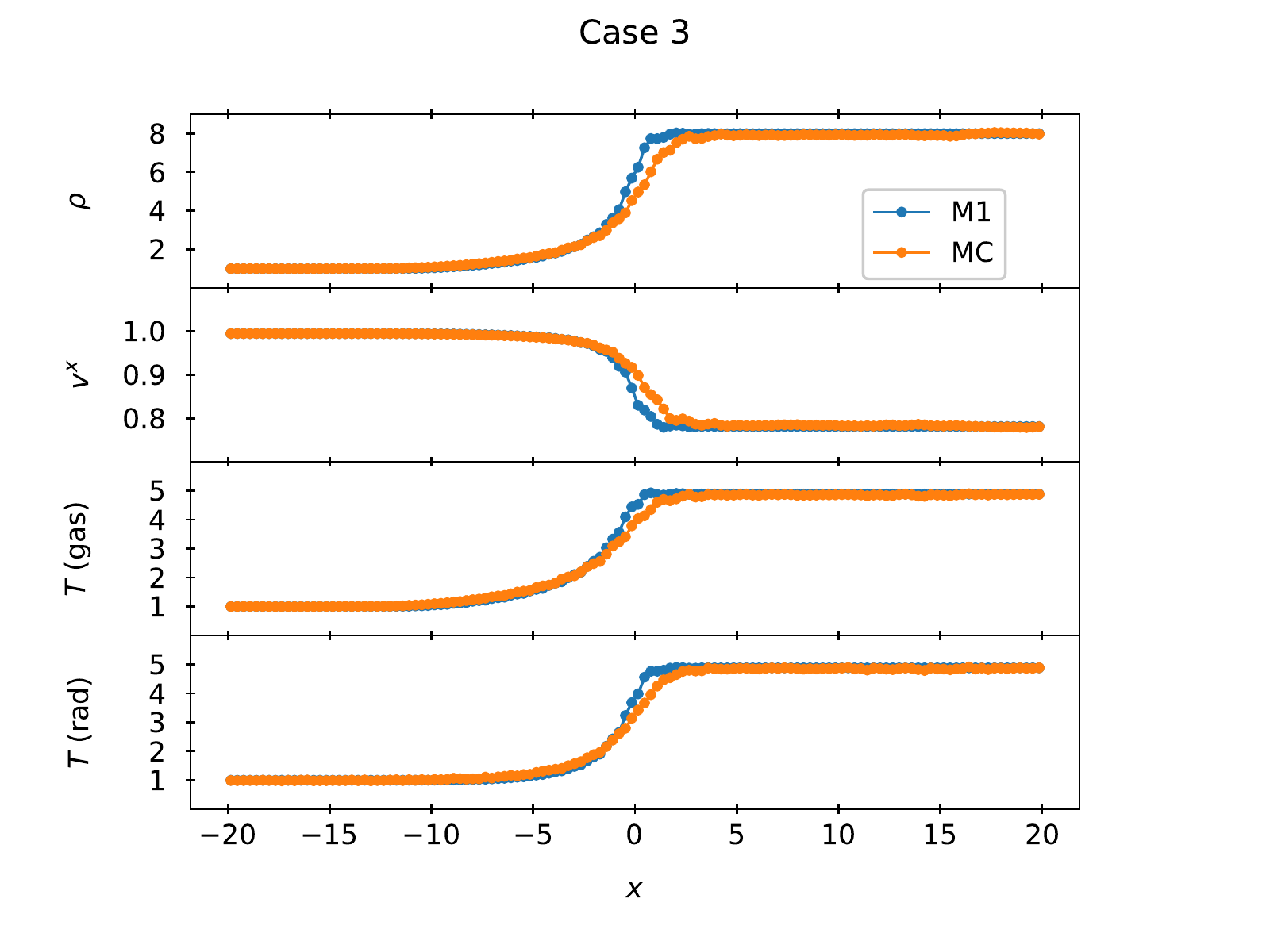} &
\includegraphics[width=0.48\textwidth]{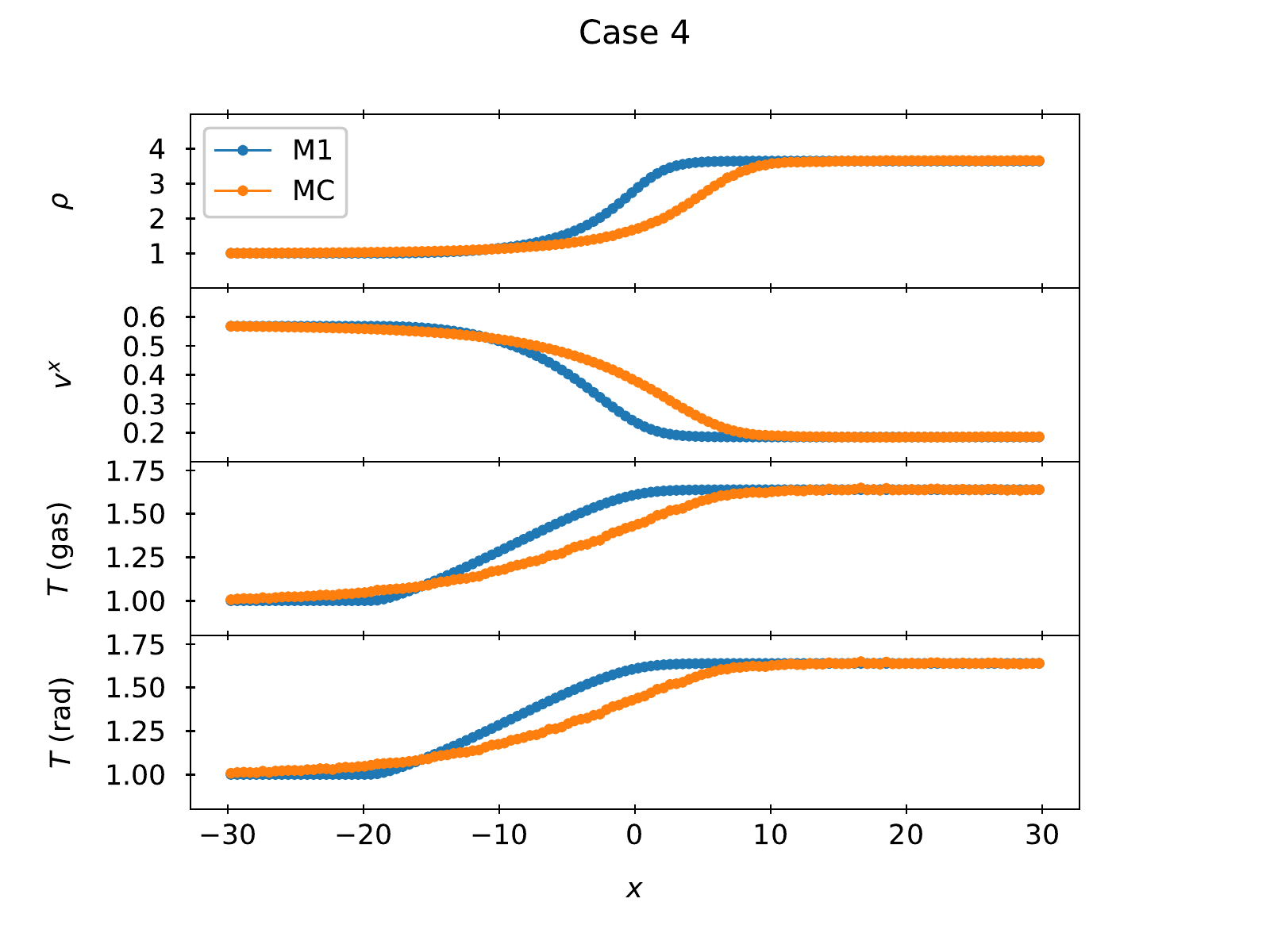} \\
(c) & (d) \\ \\
\multicolumn{2}{c}{\includegraphics[width=0.45\textwidth]{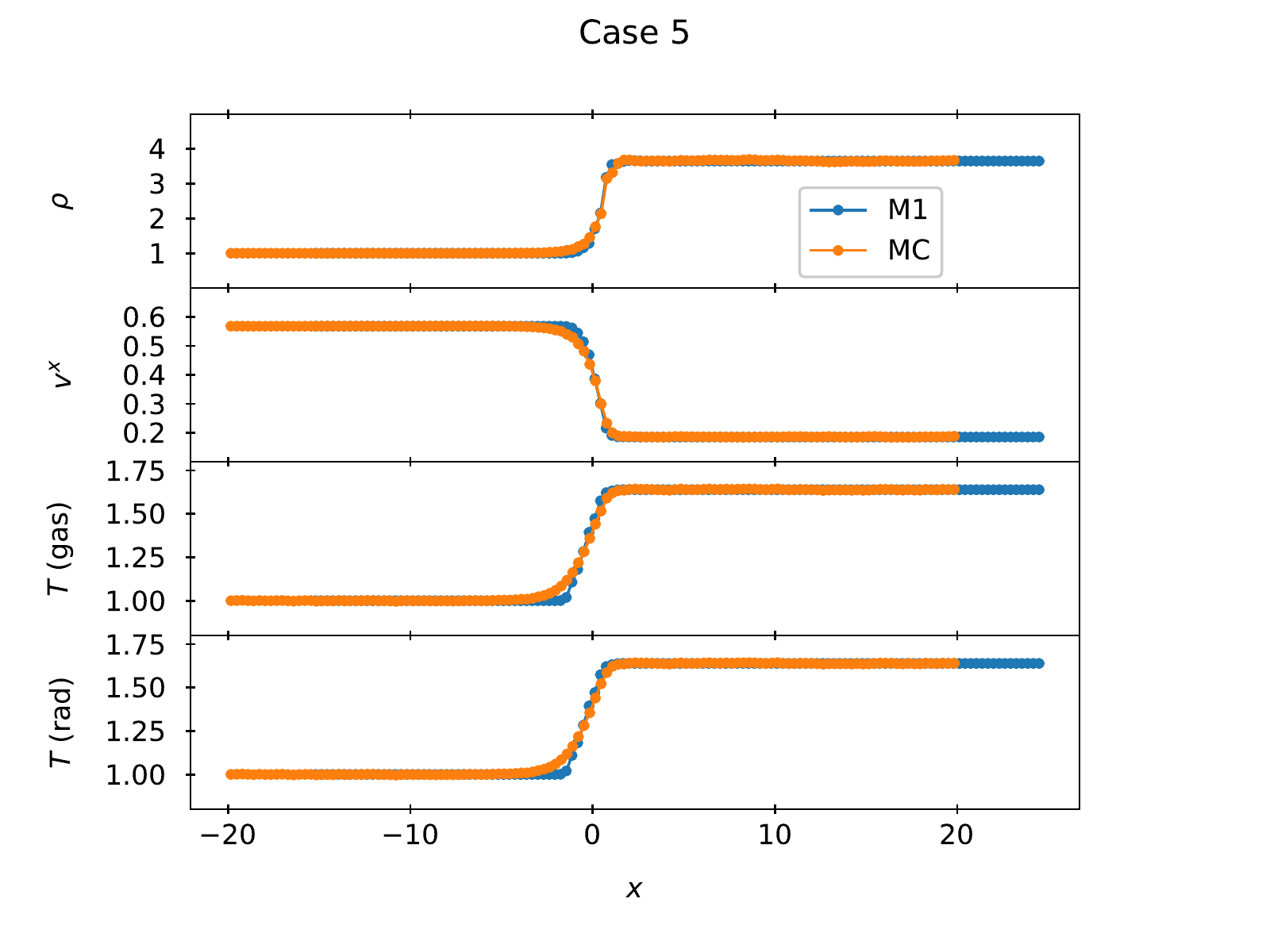} } \\ 
\multicolumn{2}{c}{(e)}
\end{tabular}
\caption{Radiating shock tube tests comparing our IMC solutions (in orange) against our M1 results (in blue).
Units are chosen so that $c = 1$, and all gas quantities other than velocity are normalized to their far-upstream values. The cases depicted in panels (a) through (c) are gas-pressure dominated throughout, and contain flows which are non-relativistic, mildly relativistic, and highly relativistic, respectively. The case shown in panel (d) is radiation-pressure dominated and relativistic. The case in panel (e) has higher radiative opacity but is otherwise identical to (d).} 
\label{fig:RadTube}
\end{figure*}

Figure~\ref{fig:RadTube} compares the M1 and IMC solutions for density ($\rho$), velocity ($v^x$), gas temperature, and radiation temperature as a function of position. We find the agreement between IMC and M1 solutions is excellent in cases 1, 2 and 5. Disagreements are most notable in cases 3 and 4, with case 4 being the most obvious. 
The extent of the disagreement has a strong dependence on the optical depth across the domain of the problem. For example, case 4 has the most disagreement and also the lowest optical depth ($\tau = 5.8$ across the downstream half of the domain).  
By comparison, case 5, which is the same as case 4 except for a higher opacity ($\tau = 51$ across the downstream half of the domain), 
shows much better agreement between the two methods. This is expected behavior, since by design the M1 closure will 
faithfully solve the radiative transfer equation in the optically thick limit, but makes assumptions which are only 
approximately correct in the marginally optically thick regime. 

\begin{figure}[htb!]
  \begin{center}
    \includegraphics[width=0.75\textwidth]{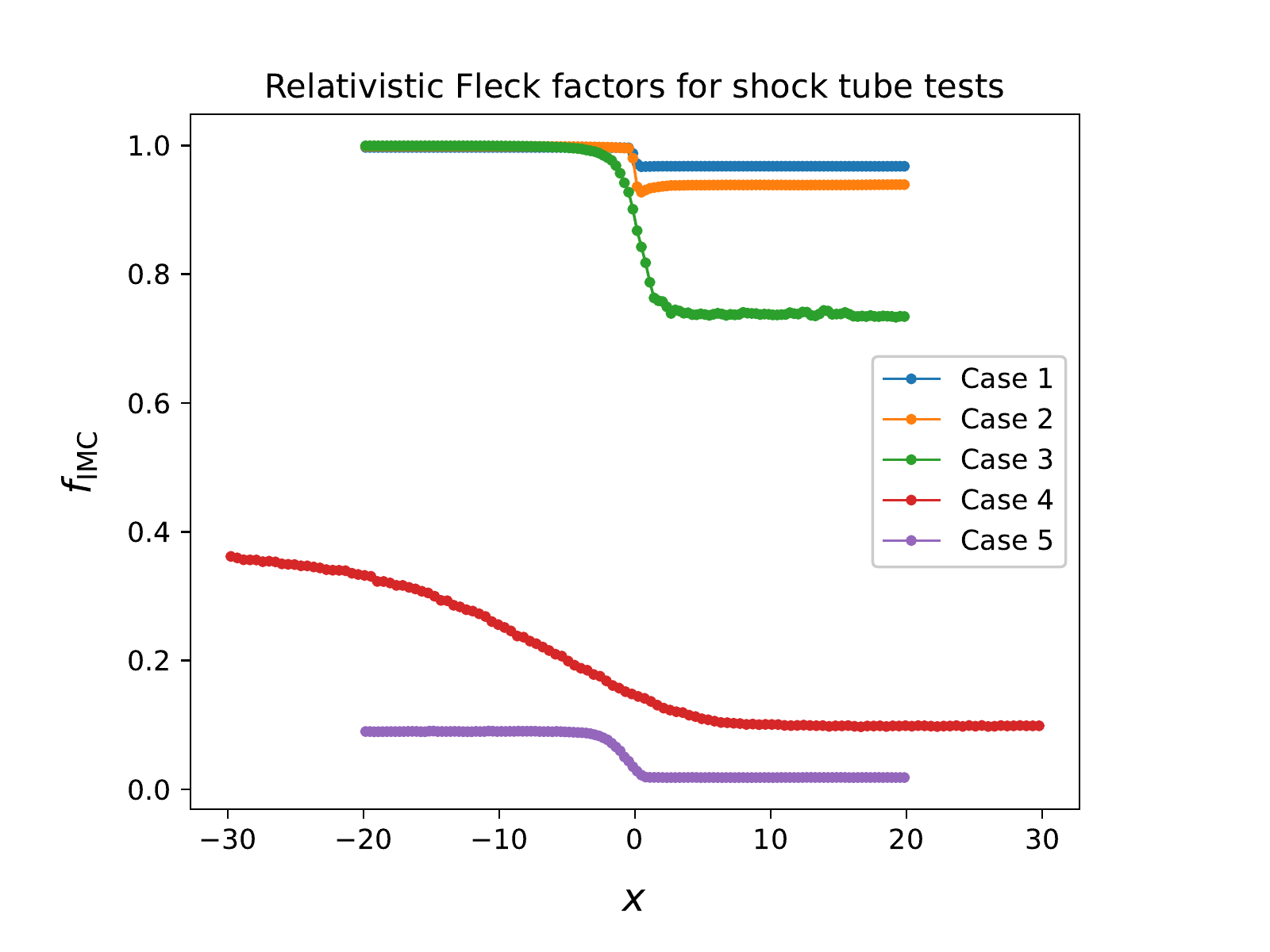}
  \end{center}
  \caption{Fleck factors $f_{\rm IMC}$ for all five radiating shock tests. For cases 3 through 5, $f_{\rm IMC}$ is small enough that these problems must be solved implicitly (with $\alpha=1$) to maintain stability at hydrodynamic timescales.}
  \label{fig:FleckFactorsRadTube}
\end{figure}

Figure \ref{fig:FleckFactorsRadTube} shows how the Fleck factor $f_{\rm IMC}$ varies across the domain once the steady-state solution has been reached in all five cases. The Fleck $\alpha$ parameter was set to 1.0 in all five cases. The lowest values of $f_{\rm IMC}$ occur in case 5, where they become less than 0.02 on the downstream side of the shock, indicating that we are taking time steps at least 10 times longer than we would be able to if we did not use a Fleck factor. We have also experimented with changing the Fleck $\alpha$ parameter for case 5, using $\alpha = 0.5$. Despite some early transitory differences at the shock position, the solutions very quickly converge within a few cycles, so the value of $alpha$ has no noticeable impact on the steady-state values of the quantities plotted in figure~\ref{fig:RadTube}.

\subsection{Compton equilibration off relativistic electrons}
\label{sec:ComptonEqTest}

In this next test we initialize photons at a single frequency $\nu_0$ in a uniform region of gas with zero velocity in the lab frame. 
We consider gas that is fully ionized, with free electron temperature $T_e$, electron number density $n_e$, and free electrons make up a fraction $f_e$ of the total number density of particles, which is constant in our problem.
The photons have number density $n_\gamma$, and they Compton scatter off the free electrons. In the process, energy is exchanged between the gas and
radiation as the volume is kept constant, and over time they equilibrate at a temperature $T_{f}$. By construction, 
photons are neither created nor destroyed in this test, and their final distribution is described by a Wien distribution with temperature $T_{f}$ and chemical potential $\mu_{f}$. 
Whenever photon energies are specified in this section, they are measured in the comoving frame of the fluid, which is also the frame in which the electron temperature $T_e$ is defined.

This test is modeled closely on one performed in \citet{Ryan2015}, and a related test in \citet{Roth2018}. 
However, in this case we perform the test at higher $T_e$, such that the relativistic effects mentioned in section \ref{sec:ComptonMethod} become relevant when sampling electron velocities and calculating mean-free-paths. We also provide an analytic formula for the equilibrium solution $T_f$ as follows.

The mean intensity for the equilibrium Wien distribution is given by
\begin{equation}
J_\nu = \frac{2 h \nu^3}{c^2} e^{-\left( \frac{h \nu}{k T_{f}} + \mu_{f} \right)} \, \, .
\label{eq:WienJnu}
\end{equation}
Keeping in mind that the specific radiation energy density is $E_\nu = 4 \pi J_\nu / c$, the above distribution can be integrated over frequency to find the radiation energy density $E_f$ which corresponds to $T_{f}$ and $\mu_{f}$, yielding
\begin{equation}
E_f = 48 \pi \frac{ \left(k T_f \right)^4}{h^3 c^3} e^{-\mu_f} \, \, .
\label{eq:WienEnergyDensity}
\end{equation}
The photon number density is found by performing a similar integral, after having divided by $h \nu$ in the integrand, resulting in 
\begin{equation}
n_\gamma = 16 \pi \frac{ \left(k T_f \right)^3}{h^3 c^3} e^{-\mu_f} \, \, . 
\label{eq:WienPhotonNumber}
\end{equation}

To solve for the two unknowns $T_f$ and $\mu_f$, we consider two constraints: conservation of photon number and conservation of total energy in the 
combined gas-plus-radiation system. Since the volume is kept constant, 
the two conditions can be expressed as conservation of number density and energy density. We exploit the fact that the two systems equilibrate to $T_f$ and the photons equilibrate to the Wien distribution in the manner previously described. This means that conservation of photon number density is already expressed in equation (\ref{eq:WienPhotonNumber}). The total energy conservation constraint can be written as
\begin{equation}
   n_\gamma h \nu_0 + \frac{1}{\gamma_{\rm ad} - 1} \frac{n_e}{f_e} k T_i =  
   48 \pi \frac{ \left(k T \right)^4}{h^3 c^3} e^{-\mu_f} + \frac{1}{\gamma_{\rm ad} - 1} \frac{n_e}{f_e} k T_f \,\, ,
\label{eq:ComptonEnergyConservation}
\end{equation}
where we have assumed an ideal gas equation of state for the gas, with adiabatic index $\gamma_{\rm ad}$, and that electrons and ions have equilibrated at temperature $T_f$. Solving the two equations we obtain
\begin{align}
\label{eq:WienTFinal}
T_f &= \frac{h \nu_0}{k} \frac{\frac{k T_i}{h \nu_0} + \frac{n_\gamma}{n_e} f_e (\gamma_{\rm ad} - 1) }{1 + 3 \frac{n_\gamma}{n_e}f_e (\gamma_{\rm ad} - 1) } ~,  \\
\mu_f &= -\ln \left[ \left( \frac{h }{k T_f} \right)^{3} \frac{c^3 n_\gamma}{16 \pi} \right] ~.
\label{eq:WienMuFinal}
\end{align}
Combining equations (\ref{eq:WienEnergyDensity}) and (\ref{eq:WienMuFinal}) we find that theradiation energy density in the equilibrium distribution is
\begin{equation}
E_f = 3 \, k T_f n_\gamma   ~,
\label{eq:WienEnergyTRelation}
\end{equation}
This implies that the mean photon energy in the equilibrium distribution is $3 \, k T_{f}$.

\begin{figure*}[htb!]
\begin{tabular}{cc}
\includegraphics[width=0.45\textwidth]{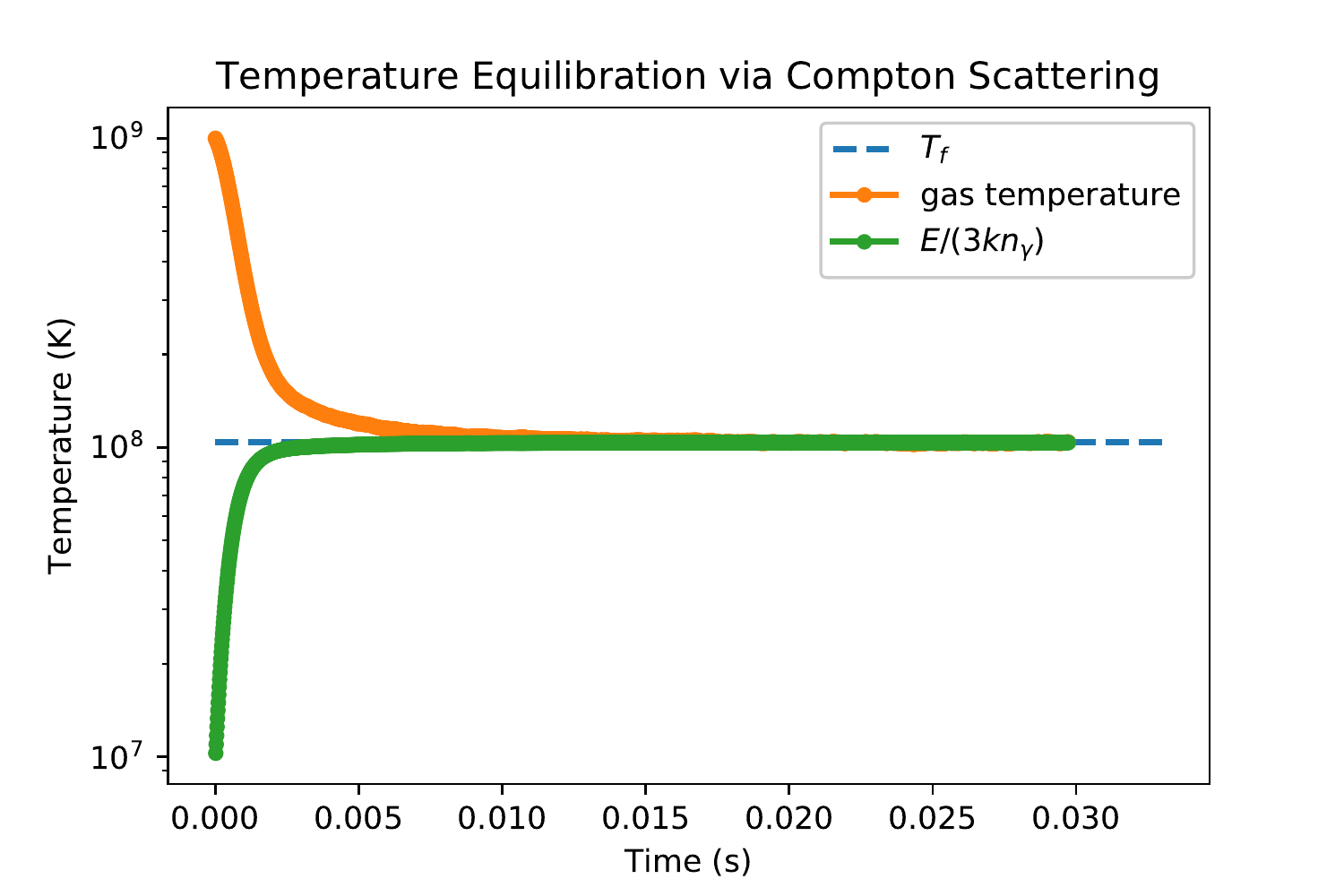} &  
\includegraphics[width=0.45\textwidth]{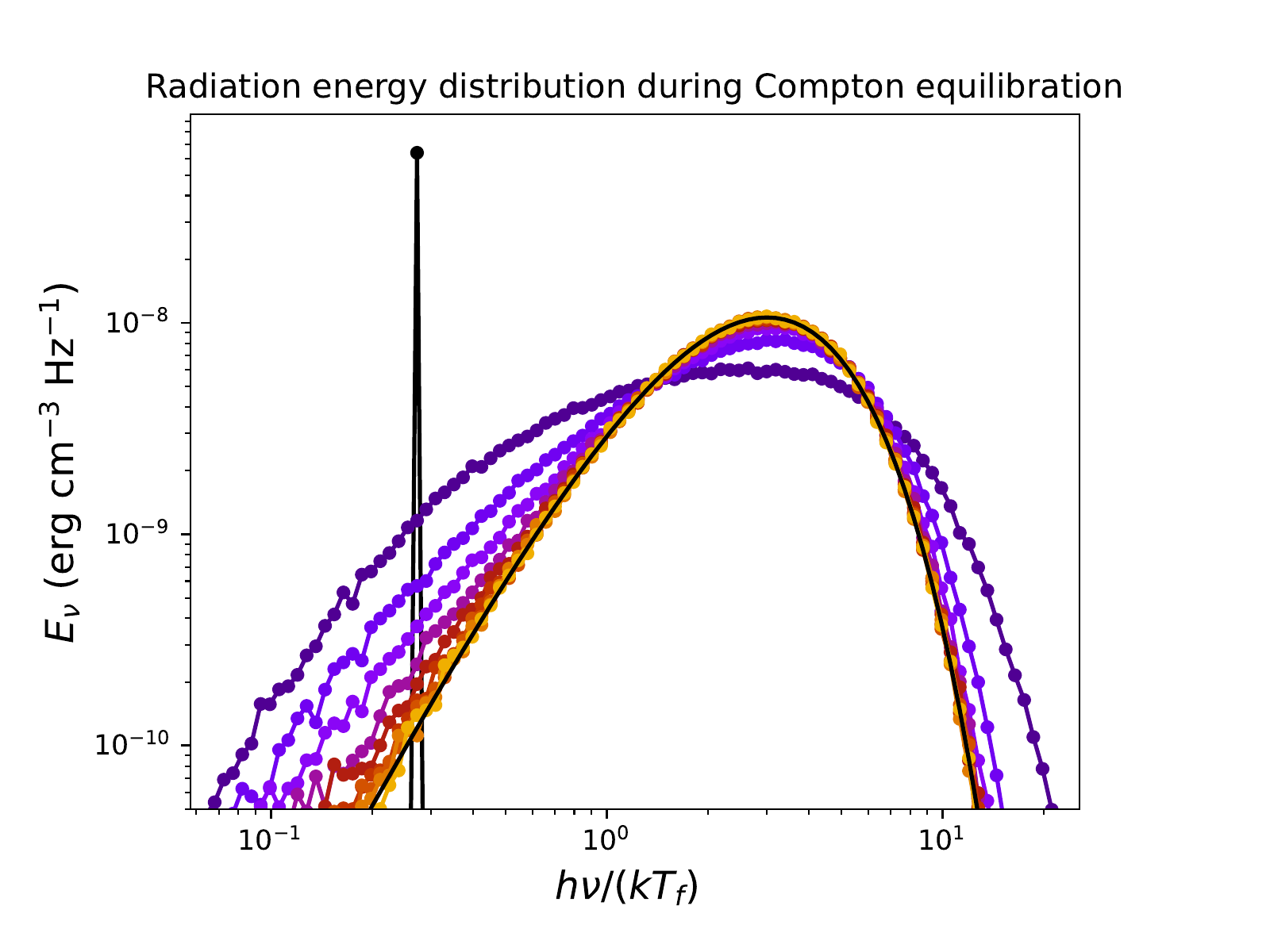} \\
(a) & (b) \\
\end{tabular}
\caption{Panel (a): Gas and radiation temperature over time in the photon-conserving relativistic Compton equilibration test. The dashed blue line on the left panel represents the equilibrium temperature. Panel (b): Photon energy distribution over time in the same test. 
The equilibrium Wien spectrum is shown as a black curve. Energy spectra correspond to time intervals of 0.003 seconds,
beginning at $t=0$ with all photon energies concentrated near $h \nu / k T_f = 0.028$, and ending at $t= 0.03$ seconds (yellow curve).}
\label{fig:ComptonEqTest}
\end{figure*}

Figure~\ref{fig:ComptonEqTest} shows results from this test with parameters 
$T_i = 10^9$ Kelvin, $n_e = 2.5 \times 10^{17}$ cm$^{-3}$, $n_\gamma = 2.38 \times 10^{18}$ cm$^{-3}$, $\nu_0 = 6 \times 10^{17}$ Hz, $f_e = 0.5$, and $\gamma = 5/3$. Equation~(\ref{eq:WienTFinal}) then gives $T_f = 1.04 \times 10^8$ Kelvin, and equation~(\ref{eq:WienMuFinal}) gives $\mu_f = 15.9$. The left panel displays the gas temperature and radiation temperature (defined by $E/[3kn_\gamma]$), as functions of time, 
both of which settle toward the predicted value of $T_f$. The right panel shows how the spectral distribution $E_\nu$ of the radiation evolves over time as the photons repeatedly Compton scatter off hot electrons. Eventually $E_\nu$ converges to the Wien spectrum 
corresponding to $T_f$ and $\mu_f$ given by equations (\ref{eq:WienTFinal}) and (\ref{eq:WienMuFinal}).

\begin{figure}[htb!]
\includegraphics[width=0.45\textwidth]{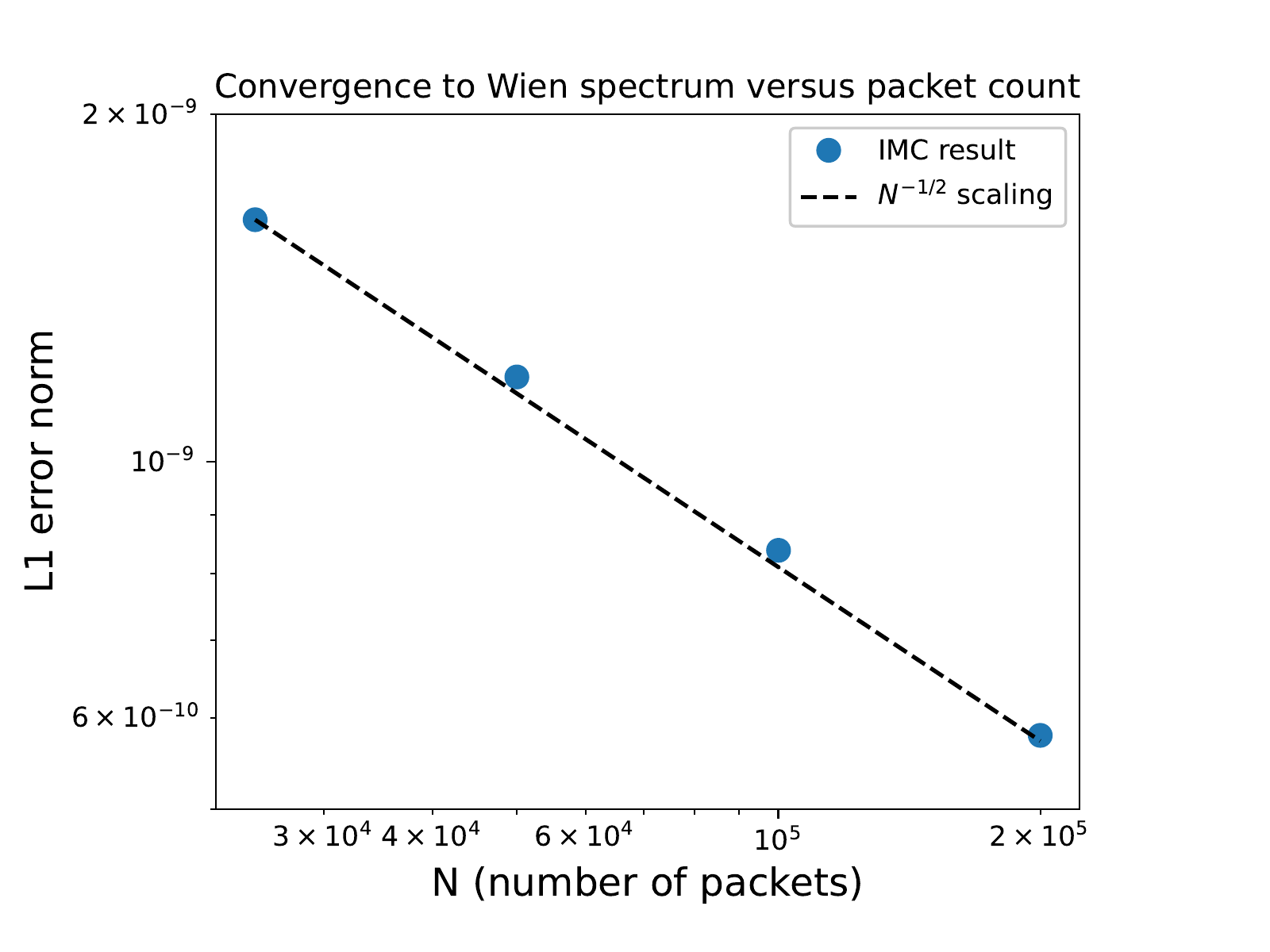}   
\caption{For the Compton equilibration test, the IMC spectra converge to the analytic Wien spectrum with an error that scales as $N^{-1/2}$ where $N$ is the number of packets.}
\label{fig:WienError}
\end{figure}

The spectra in panel (b) of Figure~\ref{fig:ComptonEqTest} used $2\times10^5$ packets. In Figure~\ref{fig:WienError}, we show how the L1 error norm for the steady-state IMC spectrum compares to the analytic Wien spectrum as a function packet number $N$. As expected we find the error scales as $N^{-1/2}$.

\subsection{Compton scattering angle distribution test}

For these tests, we again initialize photon packets at a single energy $h \nu$ and allow them to Compton scatter off a population of thermal electrons, but this time we keep track of the distribution of photon scattering angles. We are careful to consider only those scattering events where the incident photon energy is at the value we initialize --- that is, we are not describing here the effects of multiple scattering that were studied in section~\ref{sec:ComptonEqTest}.

\sloppy 
The first case we consider corresponds to the simultaneous limits $h \nu /(m_e c^2) \to 0$ and $k_B T_e /(m_e c^2) \to 0$, which is the special case of Thomson scattering, with differential scattering cross section given by 
\begin{equation}
\frac{d \sigma}{d \Omega} = \sigma_T \, \frac{3}{16 \pi} \left(1 + \cos^2\theta_s \right)  ~,
\label{eq:ThomsonDifferentialCS}
\end{equation} 
where $\sigma_T$ is the Thomson total scattering cross section and $\theta_s$ is the photon scattering angle in the rest frame of the target electron prior to scattering. We will work in terms of the probability distribution for the photon scattering 
angle $\theta_s$, $d{\cal P}/d\Omega = (1/\sigma_T) d\sigma/d\Omega$.
Introducing $\mu_s \equiv \cos\left(\theta_s\right)$, and performing the integral over the azimuthal angle, we can then write
\begin{equation}
\frac{d \cal{P}}{d \mu_s} = \frac{3}{8} \left(1 + \mu_s^2 \right)  ~.
\label{eq:dPdmuThomson}
\end{equation}

\begin{figure*}[htb!]
\begin{tabular}{cc}
\includegraphics[width=0.45\textwidth]{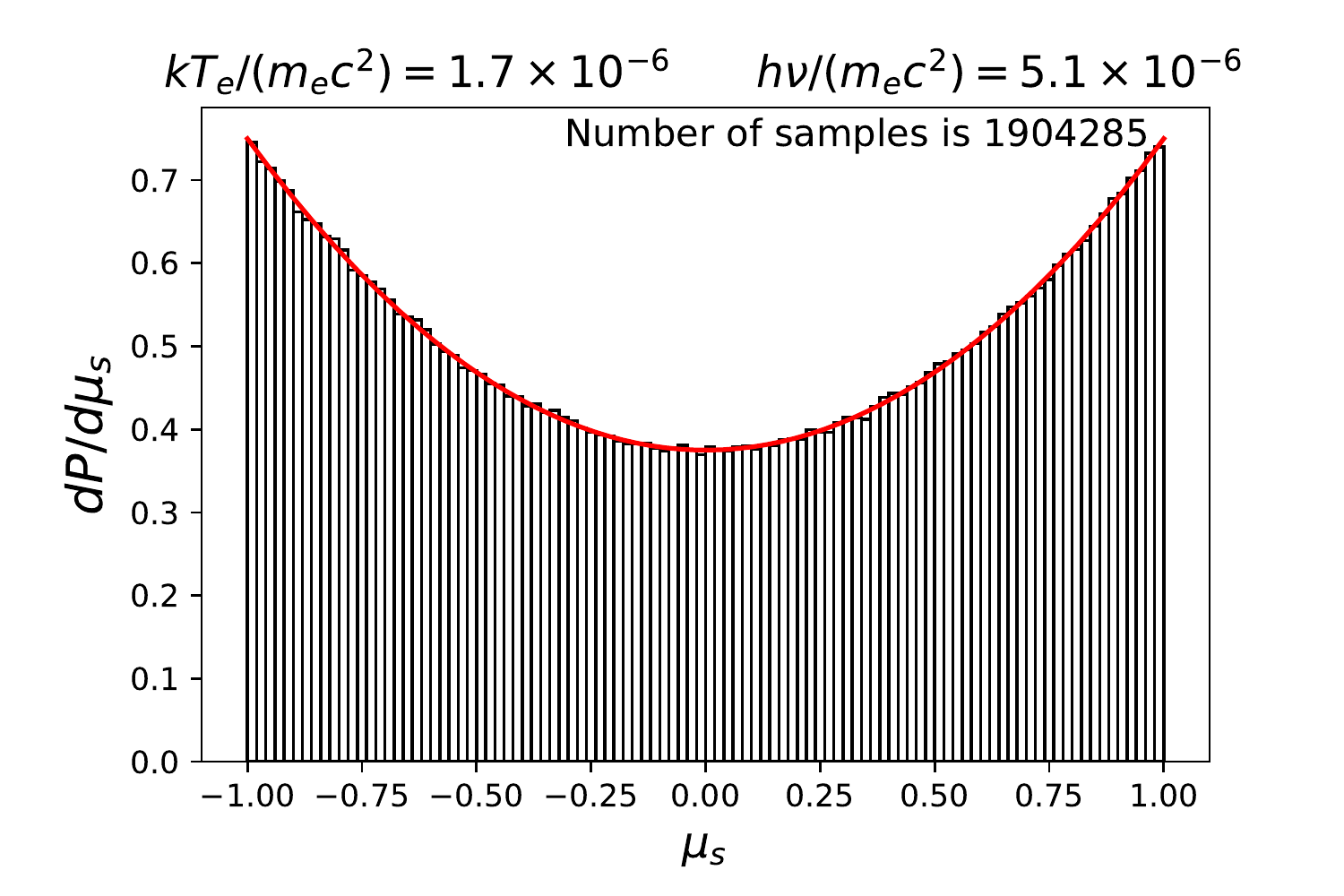} &  
\includegraphics[width=0.45\textwidth]{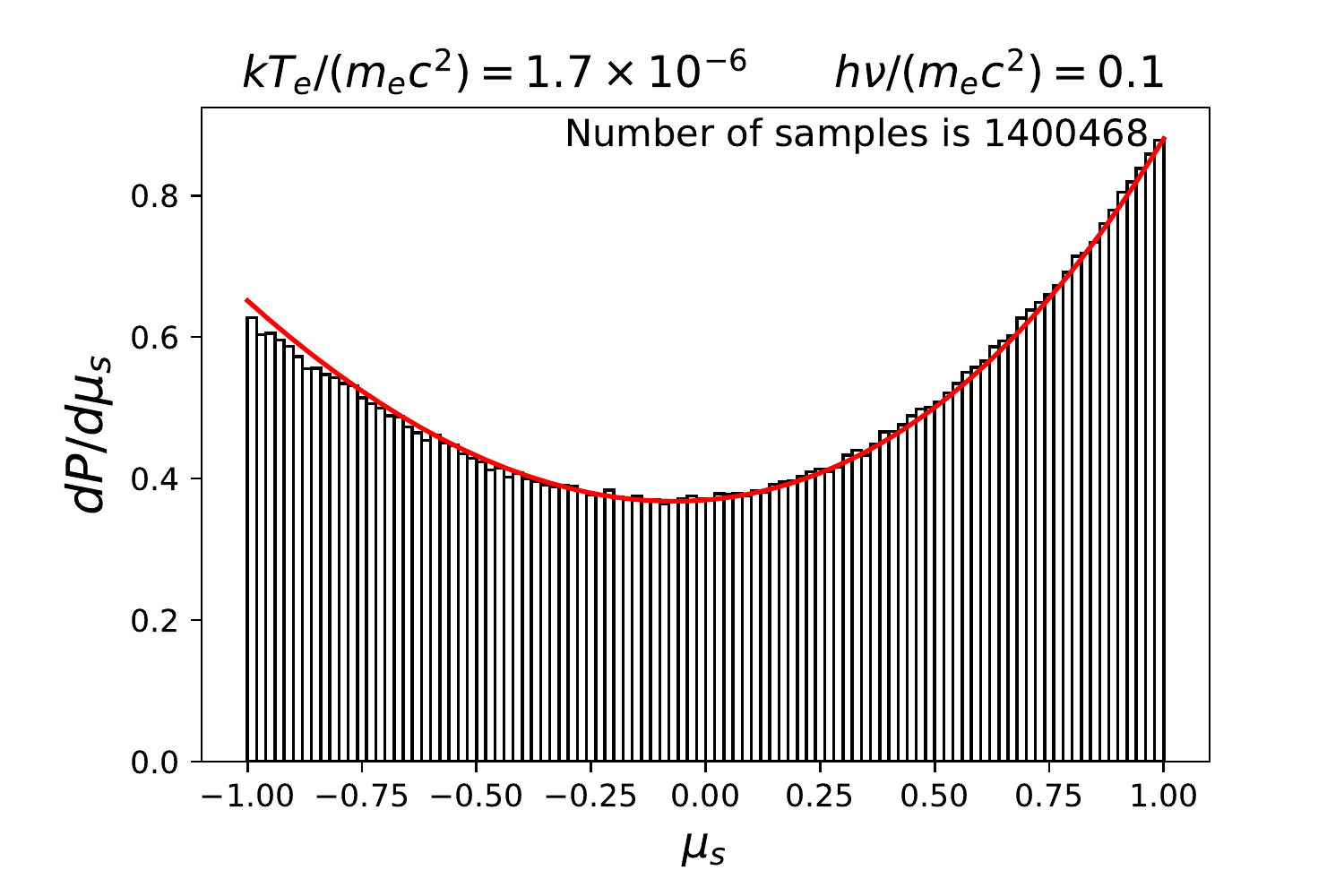} \\
(a) & (b) \\
\\
\includegraphics[width=0.45\textwidth]{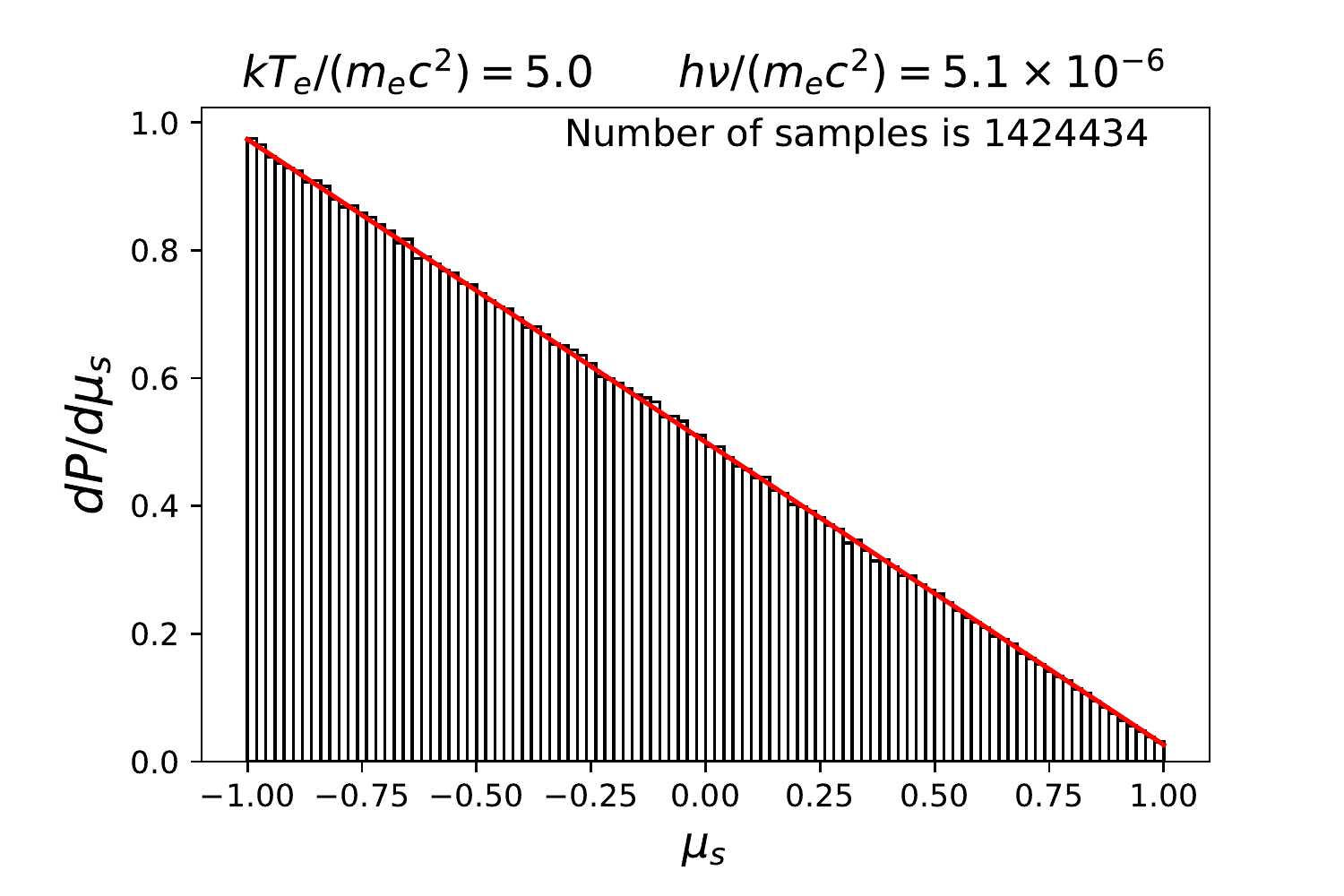} &
\includegraphics[width=0.45\textwidth]{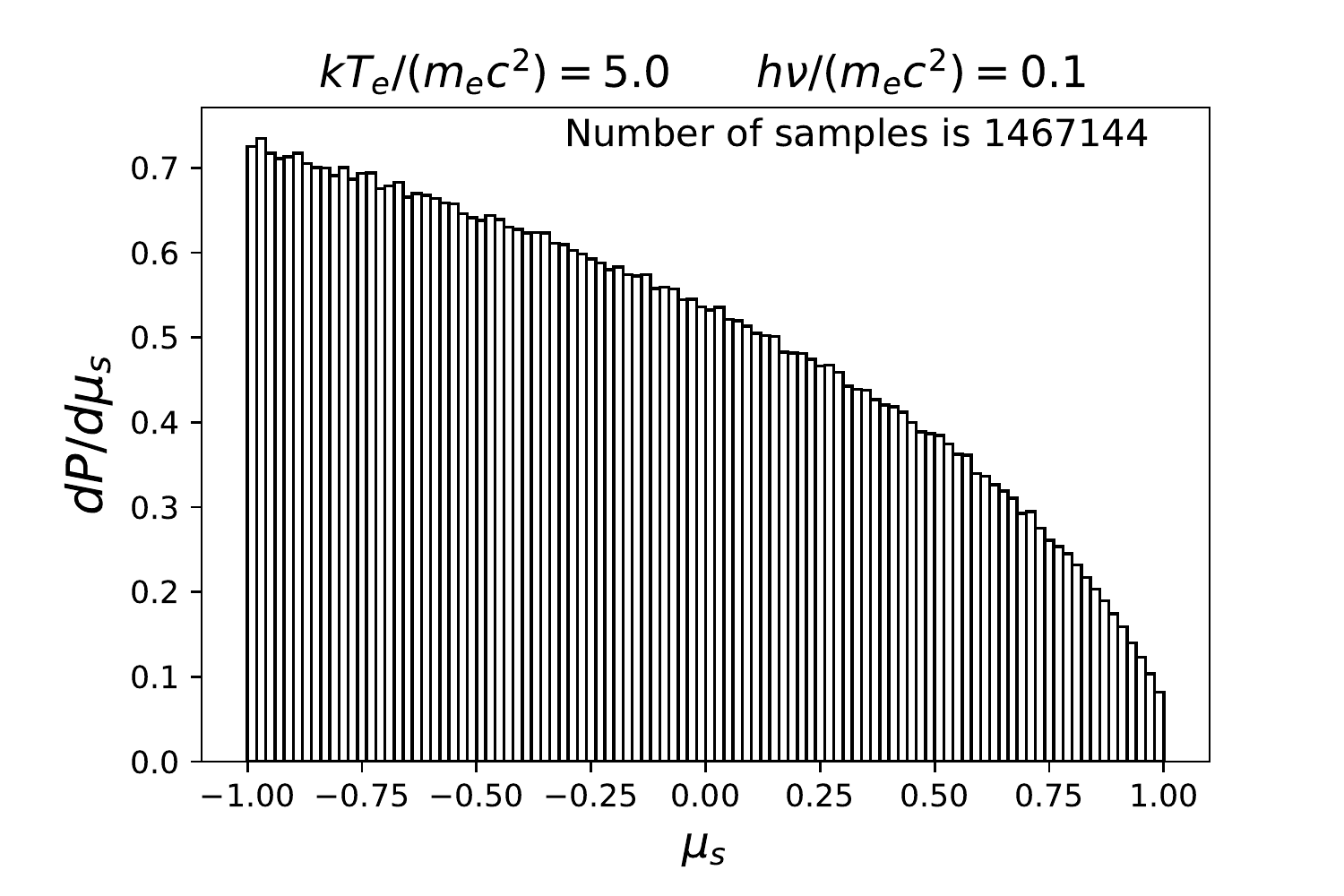} \\
(c) & (d)
\end{tabular}
\caption{Histograms of the cosine of the photon scattering angle from our Monte Carlo sampling procedure. The top and bottom rows correspond to relatively low and high temperatures, respectively, for the target population of electrons. The first and second columns correspond to relatively low and high incident photon energies, respectively. When available, we have plotted the analytic distributions from \citet{Sazonov2000-1} as red envelopes. In all cases where the approximate distribution is available, we find satisfactory agreement with our Monte Carlo results. The small discrepancy near $\mu_s = -1$ in panel (b) is expected because the analytic distribution is only first order in $h \nu / m_e c^2$.}
\label{fig:SazonovSunyaev}
\end{figure*}

In Figure~\ref{fig:SazonovSunyaev} we show histograms with randomly generated $\mu_s$ for four separate cases. 
Panel (a) of this figure corresponds to the Thomson scattering limit we have just described. Here we can compare our distribution of $\mu_s$ to equation (\ref{eq:dPdmuThomson}), and we find good agreement. 

Panel (b) of Figure~\ref{fig:SazonovSunyaev} shows the distribution of $\mu_s$ for the case of higher-energy incident photons ($h\nu = 0.1, ~m_e c^2 \approx $ 50 keV) scattering off electrons at the same temperature as before, $T_e = 10^4$ Kelvin. These higher-energy photons have a higher probability of forward-scattering ($\mu_s > 0$) than was the case in the Thomson limit. \citet{Sazonov2000-1} provide an approximate analytic formula for the scattering distribution in this regime, which is plotted in red in the figure, although we have adjusted the normalization constant for the distribution to ensure the probability integrates to unity. The slight discrepancy between the analytic formula and the Monte Carlo results 
at $\mu_s$ near -1 is expected \citep[and also appears in figure 1 of][]{Sazonov2000-1} because the formula is only first order accurate in $h \nu / (m_e c^2)$.

Panel (c) of Figure~\ref{fig:SazonovSunyaev} shows the distribution of $\mu_s$ for the case of low-energy photons ($h\nu/k_B = 3 \times 10^4 K$), but much hotter electrons ($k T_e = 5 m_e c^2$, or equivalently $T_e \approx 3 \times 10^{10}$ K). This leads to a strong weighting toward back-scattering, with a distribution in agreement with the corresponding formula from \citet{Sazonov2000-1}.

Finally, in panel (d) of Figure~\ref{fig:SazonovSunyaev}, we consider the higher energy photons, $h \nu / (m_e c^2) = 0.1$, scattering off the hotter population of electrons, $k T_e / (m_e c^2) = 5$. We are not aware of an analytic approximation for the scattering angle distribution in this case. Even though these photons were high enough energy that they had a tendency to forward-scatter off the colder population of electrons, when the electrons are this hot, the distribution is still weighted toward back-scattering.

\subsection{Gravitational and Doppler redshifts}

The accreting and radiating sphere problem considered by \citet{muller10} (see also \citet{oconnor15}) provides a good test of 
frequency redshifts experienced by photons traveling through gravitational fields and moving fluids. We adapt this test to the
Cartesian Kerr-Schild spacetime metric, modeling the gravitational field of a single non-rotating $10^3$ solar mass black hole
on a $N\times40\times40$ grid, where we vary the number of zones $N$ along the $x$-axis. The sides of the grid have total lengths $L_x = 70$ and $L_y=L_z=L_x/10$ in mass units.
We do not simulate the emission of thermal photons from a stellar surface as previous applications of this test
\citep{Anninos2020}, but instead launch a steady stream of photon packets from $x/R_g = 2.4$ (just outside the event horizon at $R_\text{BH} = 2 R_g$), directed along the $x$-axis, with identical 4-momenta. Absorption and scattering opacities are both set to zero in this problem, suppressing all radiation-matter interactions aside from the gravitational potential and Doppler interactions with the steady accretion flow introduced at a finite distance
from the black hole in the following manner:
\begin{equation}
v(r) =
    \begin{cases}
       0,                                                           & r  \le 8 R_\text{BH}, \\
       -0.2 c\left(\frac{r - 8 R_\text{BH}}{2 R_\text{BH}}\right),  & 8 R_\text{BH} \le r \le 10 R_\text{BH},  \\
       -0.2 c\left(\frac{10 R_\text{BH}}{r}\right)^2,               & r  \ge 10 R_\text{BH} ~.
    \end{cases}
\label{eqn:dop_vel}
\end{equation}

\begin{figure}[htb!]
  \begin{center}
    \includegraphics[width=0.5\textwidth]{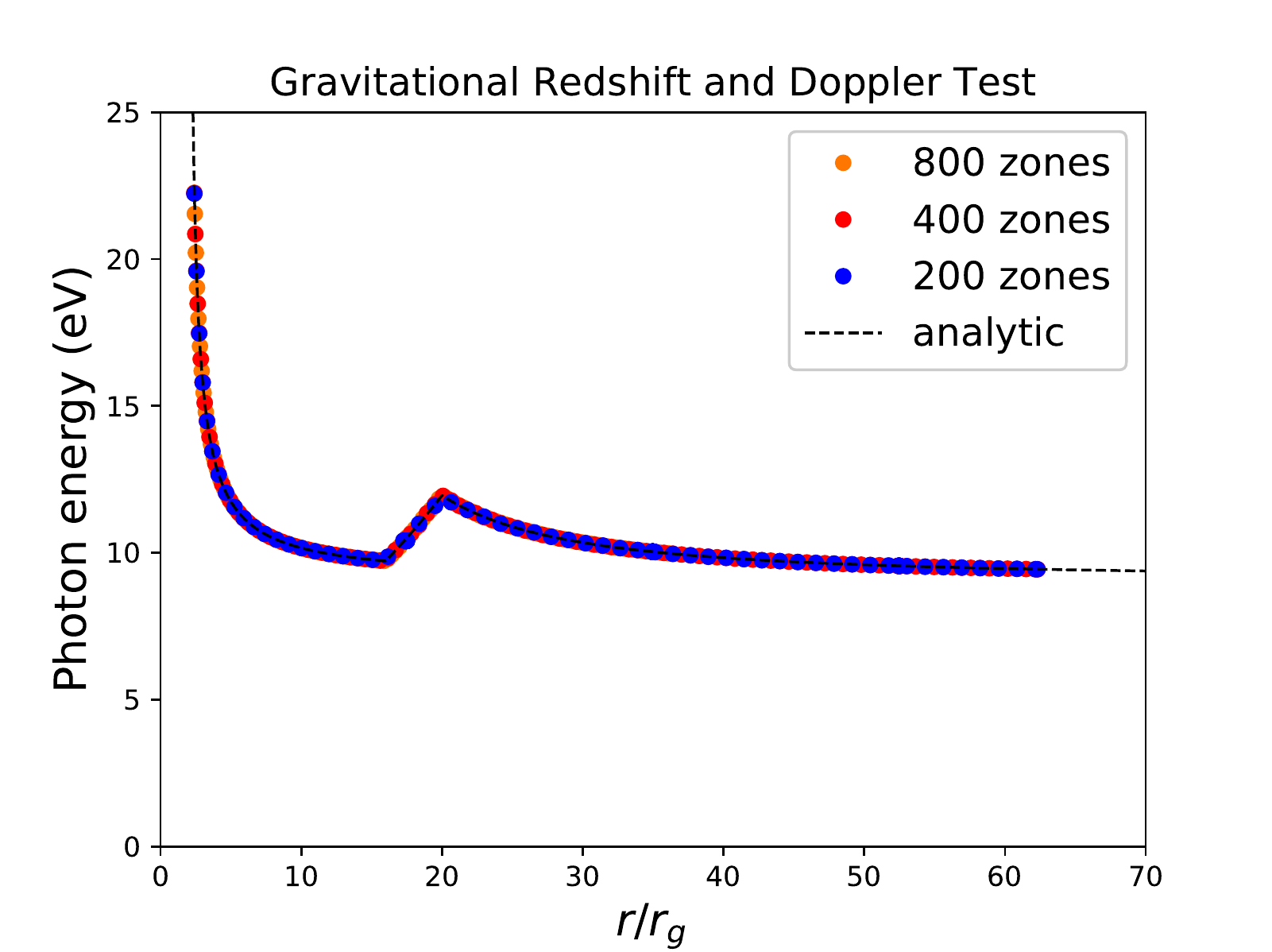}
    \end{center}
\caption{ The photon energy measured in the comoving fluid frame, $h\widehat{\nu} = -k^\alpha u_\alpha$, is plotted versus radius from the center of the black hole. Results are shown for three different grid resolutions (colored symbols), superimposed with the analytic solution (dotted line).}
\label{fig:radimcredshift}
\end{figure}

Figure \ref{fig:radimcredshift} plots photon energy measured in the comoving fluid frame, $h\widehat{\nu} = -k^\alpha u_\alpha$, as a function
of radius from the black hole. Results are shown at time $t=80$ in units with $M_{BH}=G=c=1$.
As expected, we observe photon energies to decrease with radius due to general relativistic time dilation effects,
and then increase as they propagate through the accreting fluid due to relativistic Doppler effects.

To test the accuracy of our computation of gravitational redshift and relativistic Doppler effects, we derive an exact, analytic solution to use for comparison. In the limit that the angular momentum of a black hole approaches zero, we define a radial coordinate from the Cartesian Kerr-Schild coordinates, and for purely
radial trajectories the line element can be written
\begin{equation}
\label{eq:CKSLineElement}
ds^2 = -\left(1 -\frac{R_\text{BH}}{r}\right) dt^2 + \left(1 +\frac{R_\text{BH}}{r}\right) dr^2 + \frac{R_\text{BH}}{r} (dt \, dr + dr \, dt ) ~.
\end{equation}
This implies the existence of a Killing vector field with components $K^0 = 1$ and $K^r = 0$, and a conserved quantity which we denote as $E_{\infty}$ obeying the relation
\begin{equation}
\label{eq:Einf}
E_\infty  = \left(1 - \frac{R_\text{BH}}{r}\right) k^0 - \frac{R_\text{BH}}{r} k^r  ~.
\end{equation}
Combining the above equation with the null condition for $k^\alpha$
\begin{equation}
\label{eq:NullCondition}
k^r = k^0 \left(\frac{1 - \frac{R_\text{BH}}{r}}{1 + \frac{R_\text{BH}}{r}} \right)  ~,
\end{equation}
we find
\begin{equation}
\label{eq:GravRedshiftAnalytic}
E_\infty =  k^0 \left[1 - \frac{R_\text{BH}}{r}\left(\frac{2}{1 + \frac{R_\text{BH}}{r}}\right) \right] ~.
\end{equation}
If we set the value of $k^0$ at our launch radius, then we can use equation~(\ref{eq:GravRedshiftAnalytic}) to calculate $E_\infty$, and then again to find $k^0$ as a function of radius.
Finally, if $u^\alpha$ is the fluid 4-velocity, we compute $u^\alpha k_\alpha$ by using the metric corresponding to equation~(\ref{eq:CKSLineElement}) to lower the index of $k^\alpha$.

\begin{figure*}[htb!]
\begin{tabular}{cc}
\includegraphics[width=0.45\textwidth]{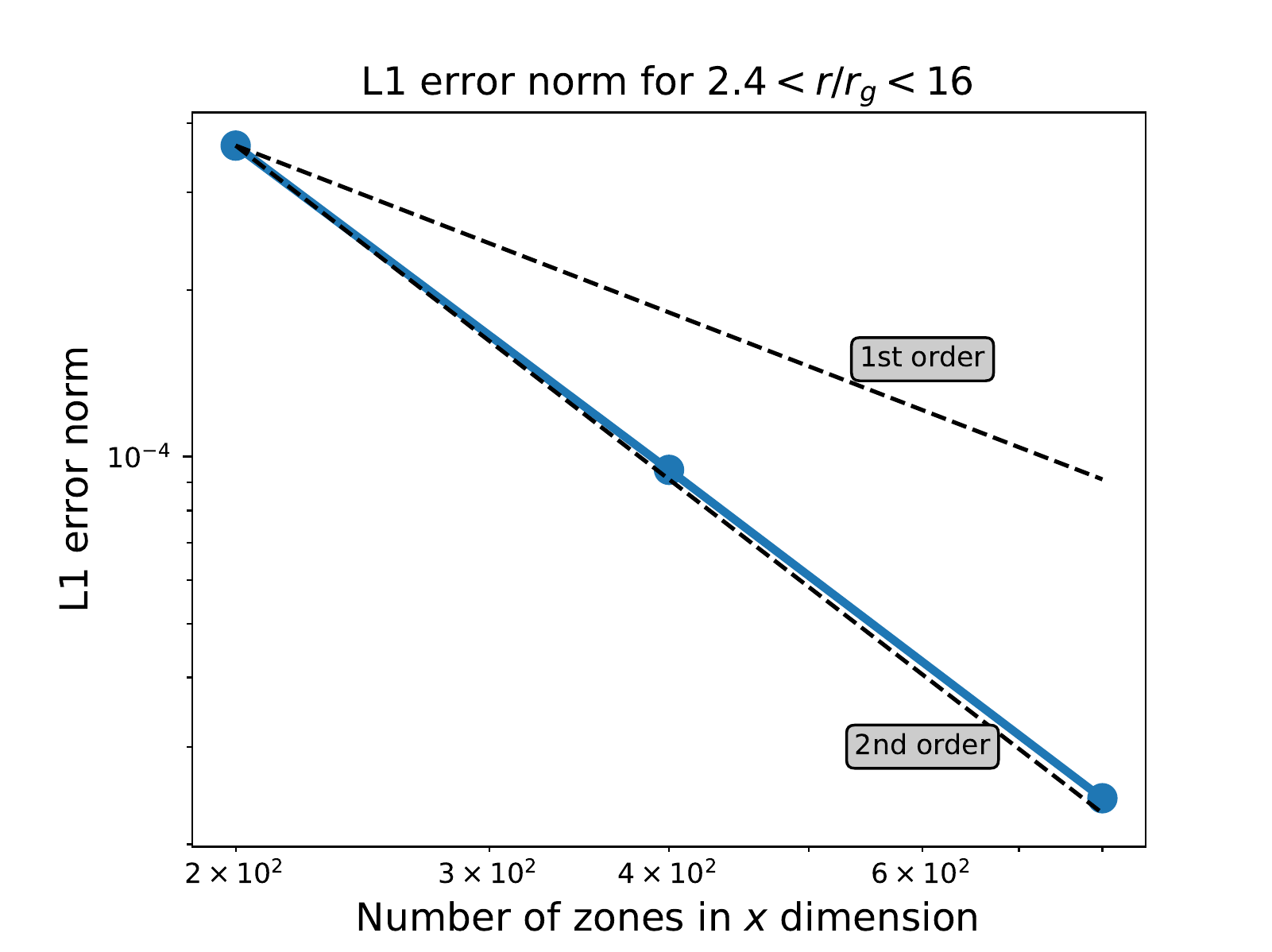} &  
\includegraphics[width=0.45\textwidth]{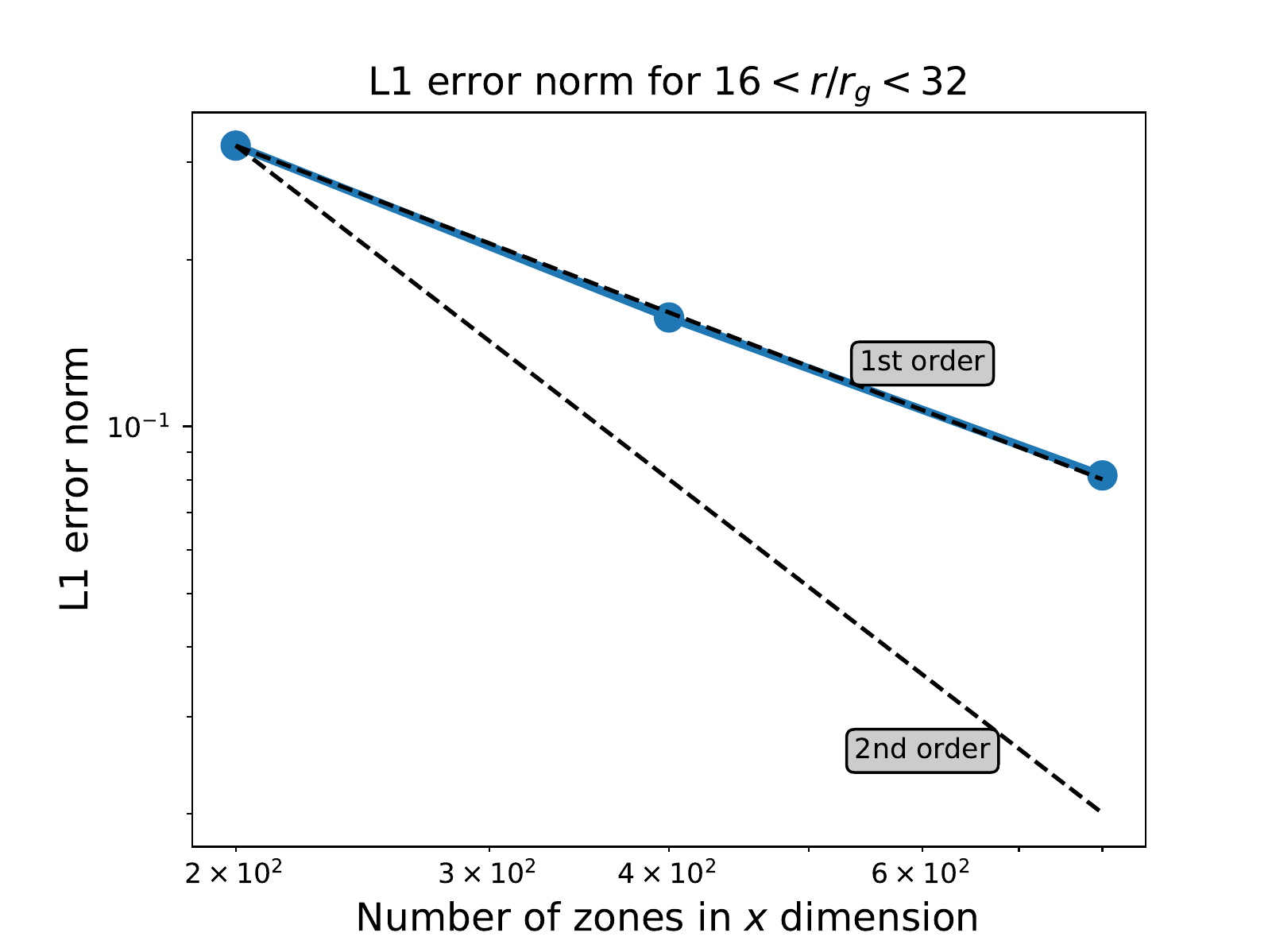} \\
(a) & (b)\\
\end{tabular}
\caption{Numerical convergence for the gravitational redshift and relativistic Doppler test} 
\label{fig:radimcredshiftError}
\end{figure*}

This analytic solution is also plotted in figure~\ref{fig:radimcredshift} as a dotted line. Even for the coarsest resolution used in that figure, the numerical solution remains within 1\% of the analytic solution across the entire domain. In figure~\ref{fig:radimcredshiftError} we study the numerical convergence of the solution as a function of spatial resolution along the $x$-axis. In panel $a$ we show how the L1 error norm decreases with increasing spatial resolution, focusing on the interval $2.4 < r/R_g < 16 $. In this region, the fluid velocity is zero, so $u^\alpha k_\alpha$ entirely reflects the gravitational redshift, which is obtained from the geodesic integration of the photon 4-momentum. The numerical solution converges at second order accuracy, as expected from the Verlet integrator used for the geodesics. In panel $b$, we plot the L1 error norm for the region $16 < r / R_g < 32$. Here the fluid velocity takes on non-zero values as given by equation (\ref{eqn:dop_vel}). When selecting the fluid velocity components that go into $u^\alpha k_\alpha$, no interpolation of the fluid velocity is performed --- the velocity value at the center of the zone is used. This causes the numerical solution to converge at first order accuracy with respect to spatial resolution. Of course, if we consider the full range of radii including both non-zero and zero fluid velocity, the first-order error dominates.

\subsection{Geodesic light beams}

For strongly gravitating systems (e.g., black holes), it is critically important that we solve the geodesic equations
(\ref{eqn:geod1}) accurately through multi-dimensional curved spacetimes in general curvilinear coordinates.
To validate our geodesic solvers we consider a light beam test proposed by \citet{sadowski13} which
launches packets at the photon orbit radius, $r=3$ in distance units of $GM/c^2$ for a non-rotating 
3 $M_\odot$ Schwarzschild black hole, so that
photons in the center of the beam should in principle be able to orbit the black hole indefinitely.
As in the previous redshift tests, we neglect coupling between gas and radiation (by setting absorption
and scattering opacities to zero), focusing on evaluating the accuracy of geodesic path integrals
but keeping to the full IMC solver framework for computing the energy and flux.
The calculations presented here are run on a two-dimensional grid (in $r$-$\phi$ coordinates), with $40\times40$
cells covering the domain $0 \le \phi \le \pi/2$ and $r_{\rm in} \leq r \leq r_{\rm out}$, where
$r_{\rm in} = 2.5$ and $r_{\rm out}=3.5$ are the inner and outer radial boundaries, respectively. The
light beam is centered at $r=3$ and given a width of $r=0.1$.

\begin{figure}[htb!]
  \begin{center}
    \includegraphics[width=0.75\textwidth]{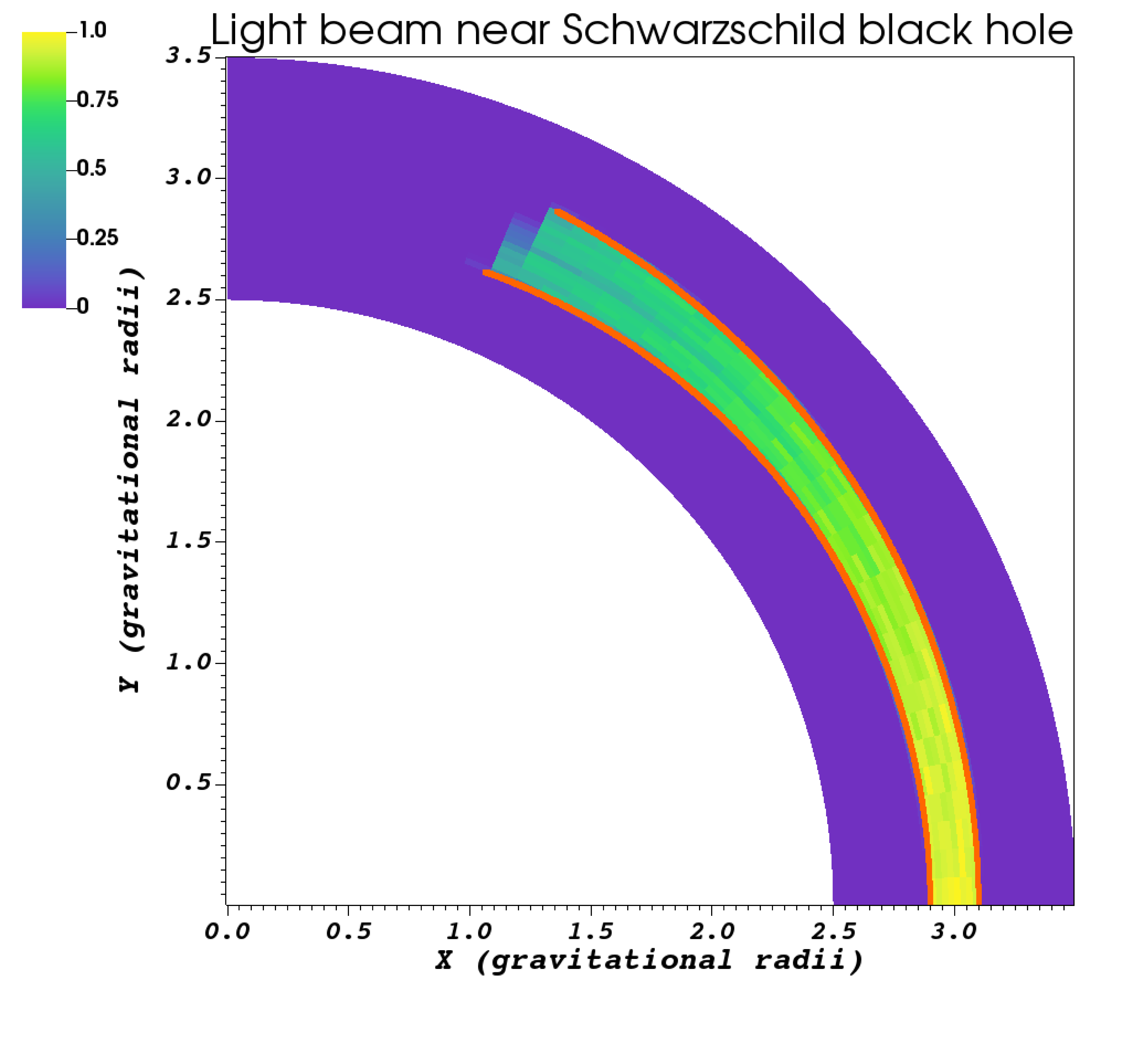}
    \end{center}
\caption{Test of 2D photon geodesic trajectories near a Schwarzschild black hole. The color scale indicates the radiation energy density as a function of its value at the injection location at $y = 0$. The orange curves represent two bounding geodesic trajectories for the beam.}
\label{fig:lightbeam}
\end{figure}

Figure~\ref{fig:lightbeam} plots the trajectory of the light beam at the final simulation time. As expected the center of the beam remains on the photon orbit radius, and widens slightly at the edges (falling towards the black hole along the inner edge, and pulling away at the outer).

We have also performed a number of single photon orbital trajectory tests through static and rotating black hole
spacetimes, evaluating errors in energy, angular momentum, Carter, and null condition constants of motion.
These tests were carried out on three-dimensional grids using the Boyer-Lindquist metric with a black hole mass
of $10^3 M_\odot$ and spins up to $a = 0.9 M_{\rm BH}$, considering both prograde and retrograde equatorial orbits,
as well as polar trajectories \citep{Teo2003}. We find, as expected, that the constants of motion are maintained
to the convergence accuracy of the solver methods: first order convergence for first order Runge-Kutta (RK1), second order
for second order Runge-Kutta and Verlet methods, and fourth order for fourth order Runge-Kutta (RK4).
A characteristic sample of parameters and results are shown in Table \ref{tab:geodesic}, where we demonstrate
convergence in the energy and Carter constants over several orbits between the RK1, Verlet, and RK4 methods. 
All calculations were run at a constant time step interval of about $0.01$ in mass units ($M = G = c = 1$).
We do not include the angular momentum or null ($k^\alpha k_\alpha = 0$) constants in the table since they evaluate to
roughly machine precision (less then $10^{-14}$ in most of the tests).

\begin{deluxetable}{ccccc}
\tablecaption{Geodesic Errors \label{tab:geodesic}}
\tablewidth{0pt}
\tablehead{
    \colhead{Orbit} & \colhead{Spin} & \colhead{Method} & \colhead{Energy constant} & \colhead{Carter constant (Q)} 
}
\startdata
equatorial, prograde   & 0.5 & RK1    & $6.1\times10^{-5}$  & $<10^{-15}$  \\
                       &     & Verlet & $6.3\times10^{-9}$  & $<10^{-15}$  \\
                       &     & RK4    & $3.4\times10^{-14}$ & $<10^{-15}$  \\
equatorial, retrograde & 0.9 & RK1    & $9.0\times10^{-5}$  & $<10^{-15}$  \\
                       &     & Verlet & $4.5\times10^{-9}$  & $<10^{-15}$  \\
                       &     & RK4    & $1.5\times10^{-14}$ & $<10^{-15}$  \\
polar, pole reaching   & 0.5 & Verlet & $7.0\times10^{-9}$  & $2.1\times10^{-8}$   \\
                       &     & RK4    & $2.2\times10^{-16}$ & $1.5\times10^{-13}$  \\
polar, maximal Q       & 0.5 & Verlet & $1.2\times10^{-8}$  & $8.4\times10^{-7}$   \\
                       &     & RK4    & $3.4\times10^{-14}$ & $1.9\times10^{-12}$
\enddata
\end{deluxetable}

To measure the relative performance of the various geodesic solvers, we repeated the 2D light-beam test using RK1, RK4, and Verlet methods, in addition to the RK2 method that was used to generate figure~\ref{fig:lightbeam}. The timing information for the calculations is reported in table~\ref{tab:geodesicPerformance}, normalized to the RK1 compute time.

\begin{deluxetable*}{ccc}
\tablecaption{Timing comparison for the light beam geodesic test \label{tab:geodesicPerformance}}
\tablewidth{0pt}
\tablehead{
    \colhead{\hspace{2cm}Method} & \colhead{\hspace{2cm}Relative compute time} & \hspace{2 cm}
}
\startdata
\hspace{2 cm}RK1    & \hspace{2 cm}1.00 \hspace{2 cm} &\hspace{2 cm}\\
\hspace{2 cm}RK2    & \hspace{2 cm}1.72 \hspace{2 cm} &\hspace{2 cm}\\
\hspace{2 cm}RK4    & \hspace{2 cm}3.15 \hspace{2 cm} &\hspace{2 cm}\\
\hspace{2 cm}Verlet & \hspace{1.85 cm}2.45 \hspace{2 cm}&\hspace{2 cm}
\enddata
\end{deluxetable*}

\subsection{Radiating Bondi accretion}
\label{sec:RadBondi}

The radiating Bondi accretion problem considered in our earlier work \citep{Fragile2012-1, Fragile2014-1} provides an excellent test
of coupled radiation-matter interactions in a hot and strongly gravitating black hole environment. We consider a variant of that test
here, comparing our current IMC results to our previous radiation treatment with M1 closure. A key difference however
is that here we work in three-dimensions on Cartesian grids using the Cartesian Kerr-Schild black hole
solution for the spacetime metric, rather than its two-dimensional spherical equivalent. Other than this difference,
the basic problem definition and parameter choices are the same, so we refer the reader to those earlier papers for 
details and merely report the salient features here.

We fix the mass of the black hole to $3 M_\odot$, the adiabatic gas index to $\Gamma=5/3$, and the gas temperature,
freefall velocity, and density to
\begin{eqnarray}
T    &=& T_0 \left({\rho}/{\rho_0} \right)^{\Gamma-1}  ~, \\
u^r  &=& -\sqrt{{2M}/{r}}  ~,  \\
\rho &=& -{\dot{M}}/{(4\pi r^2 u^r)} ~,
\end{eqnarray}
where the mass accretion rate is defined in terms of the Eddington rate $\dot{M} = x_M \dot{M}_{\rm Edd}$
with $\dot{M}_{\rm Edd} = 4\pi G M\sigma_T/(c m_p)$, and for this test we set $x_M = 0.1$.
The radiation is initialized with a temperature much smaller than the gas temperature, but
eventually increases in time as accretion heats up and thermal emissions contribute to the luminosity.
Interactions between matter and radiation are modeled with Thomson scattering ($\kappa^s$) and thermal bremsstrahlung ($\kappa^a$)
opacities assuming fully ionized hydrogen gas
\begin{eqnarray}
\kappa^s &=& 0.4 \quad \text{cm}^2 \text{g}^{-1} ~, \\
\kappa^a &=& 1.7\times10^{-25} ~T^{-7/2} ~\rho ~m_p^{-2}  \quad \text{cm}^2 \text{g}^{-1} ~. 
\end{eqnarray}

\begin{figure}[htb!]
\begin{tabular}{cc}
\includegraphics[width=0.45\textwidth]{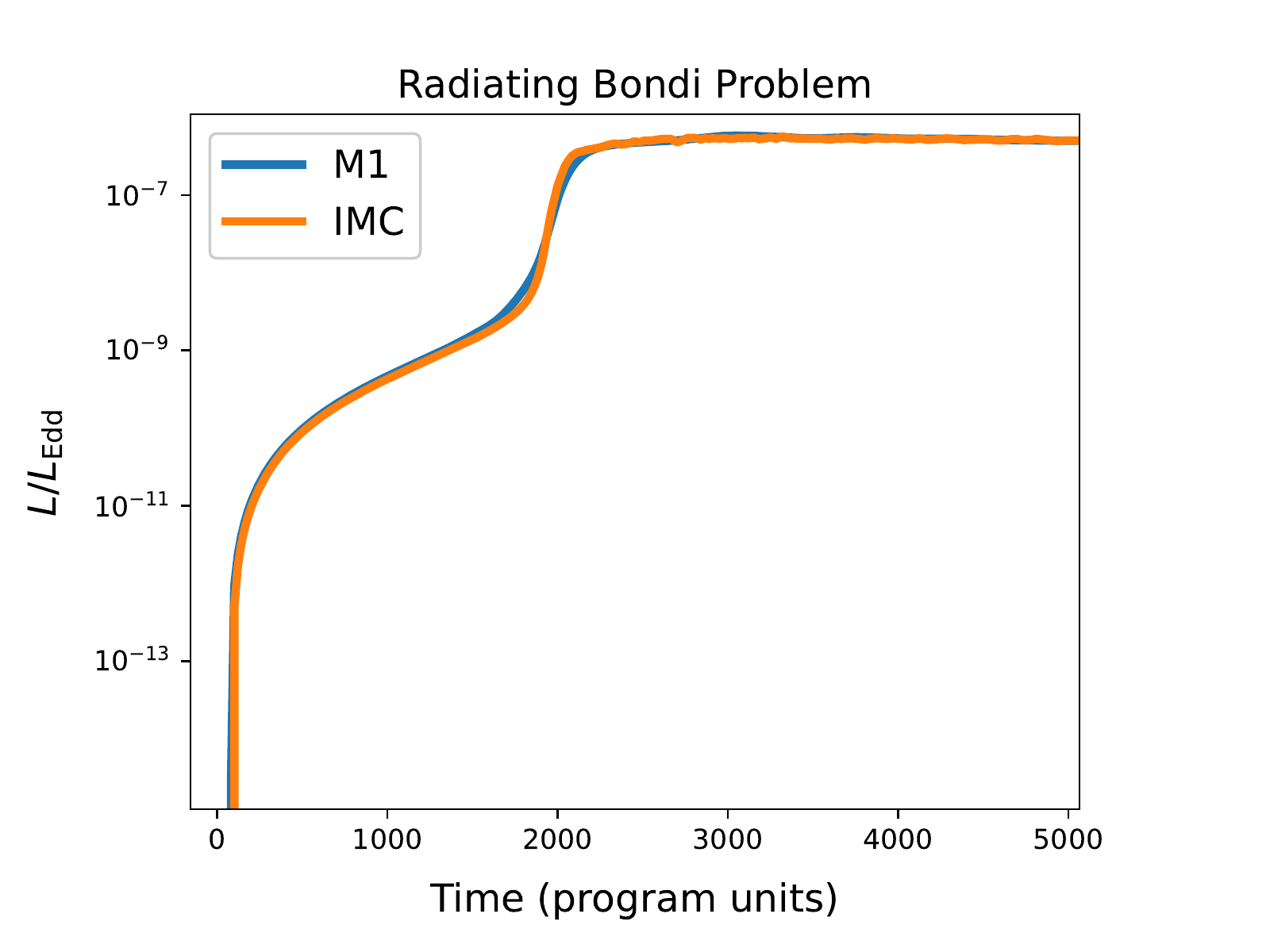} &  
\includegraphics[width=0.45\textwidth]{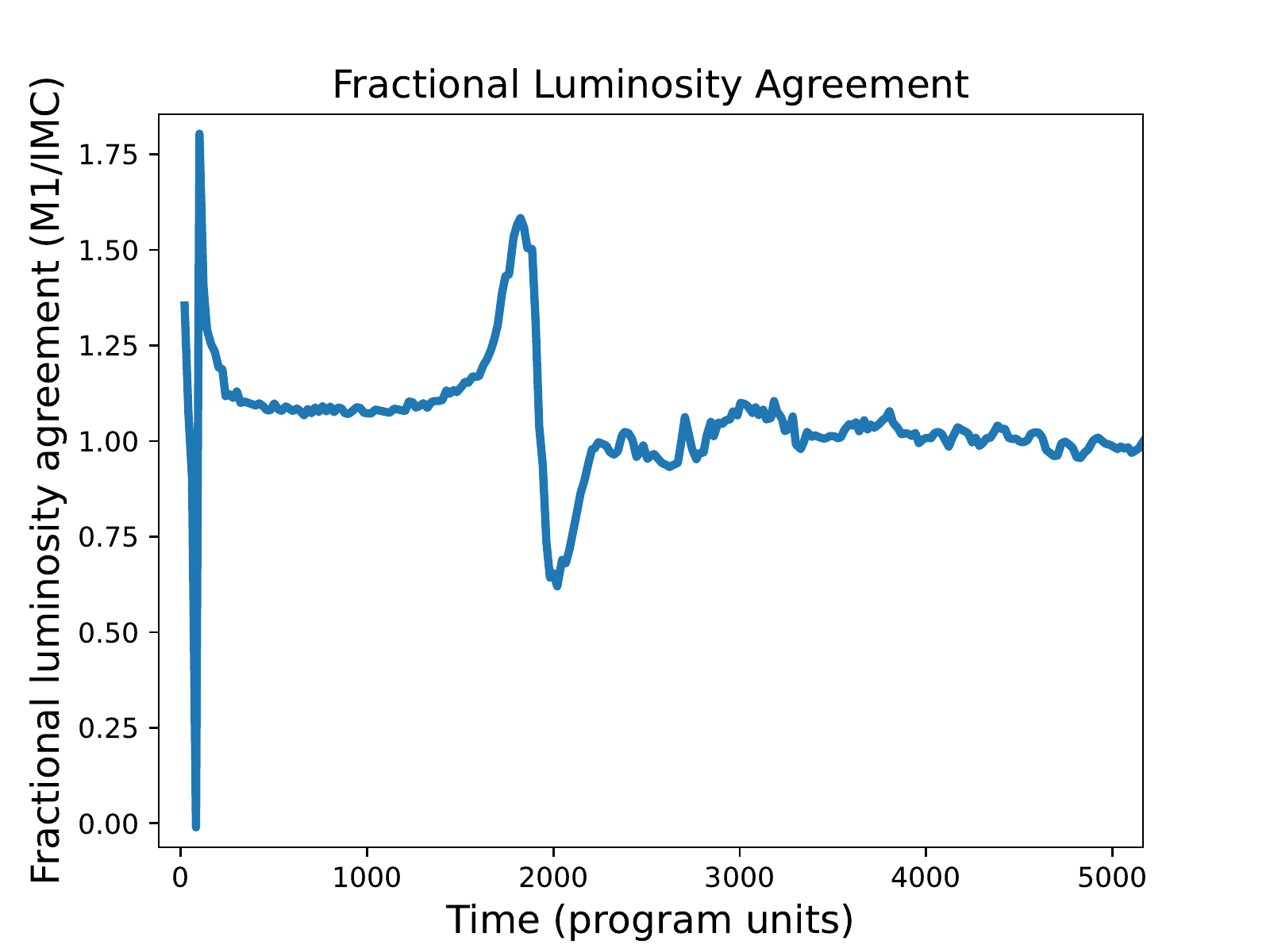} \\
(a) & (b) \\
\end{tabular}
\caption{Panel (a): Luminosity escaping through a spherical surface of radius 2000 $R_g$, versus time in program units, for both the M1 and IMC calculations. Panel (b): M1 luminosity divided by IMC luminosity versus time in program units. After a brief initial transient, between $t = 400$ and $t = 1200$, the two methods agree to within 10\%. Between $t = 1200$ and $t = 2200$ there is larger disagreement as the radiation wave from accretion near the black hole reaches the flux measurement surface. This is then followed by a period of even better agreement: between $t = 2200$ and $t = 5000$, the root-mean-squared fractional difference between the luminosities is $3.9$\%.}
\label{fig:radbondi_luminosity}
\end{figure}

The luminosity, defined as
\begin{equation}
L = \int_S \sqrt{-g} F^r dA_r ~,
\end{equation}
where $dA_r$ is the surface area element normal to the radial direction, agrees very well between the two methods (IMC versus M1). Figure \ref{fig:radbondi_luminosity} shows a comparison of the luminosity escaping through a spherical surface of radius 2000 $R_g$ centered on the black hole.

\section{Performance}
\label{sec:Performance}

A subset of the code tests presented in this article, particularly those that were validated against equivalent M1 calculations,
provide an opportunity to compare relative CPU resources required by the two different treatments within the same code framework. 
A few 1D radiating shock tube tests and the 3D radiating Bondi accretion problem are selected to showcase this comparison. 

We note our current implementation does not redistribute particles for load-balancing purposes.
Domain exchanges are performed only when particles propagate into domain-shared cells while transiting to their
neighboring partition. The following performance comparison therefore does not account for communication costs (or benefits) of
either mesh re-partitioning or particle load balancing beyond field and particle cross-over exchanges.

\begin{deluxetable*}{ccccc}
\tablecaption{Relative Performance Chart \label{tab:performance}}
\tablewidth{0pt}
\tablehead{
  \colhead{Test} & \colhead{\# processors}  & \colhead{Stop time } & \colhead{Average \#  active packets } & \colhead{IMC/M1 run time}
}
\startdata
1D-Case3 & 64 & 4.0 crossing times & 5.2e6 & 510  \\
 & & (upstream velocity / grid length) & & \\
1D-Case4 & 32 & 3.1 crossing times & 1.7e7 & 2310  \\
1D-Case5 & 32 & 1.3 crossing times & 9.9e6 & 1120  \\
3D-Bondi & 64 & 5000 mass units ($G = c = 1$) & 1.4e8 & 27 (M1 1 group)\\
 &  &  & & 5.1 (5 groups) \\
&  &  & & 2.0 (10 groups)\\
&  &  & & 0.51 (20 groups) \\
&  &  & & 0.087 (40 groups) \\
\enddata
\end{deluxetable*}

Table \ref{tab:performance} summarizes our results across four tests: the first three are 1D radiation shock tube
problems (cases 3, 4 and 5 as described in section \ref{sec:RelativisticRadiatingShocks}), and
the last is the 3D Bondi accretion test from section  \ref{sec:RadBondi}. The first column identifies specific tests, the second column indicates how many processors were used, the third column shows how long in program time units the calculations were run, the fourth column records the number of packets active across the grid (averaged over time), and the fifth column is the ratio of the wall clock time required to complete the calculation for IMC relative to M1. All of the M1 calculations for the 1D relativistic shock tests were performed with a single frequency group. For the 3D Bondi test, we included multiple timing comparisons performed with different numbers of M1 frequency groups. The opacity used in the Bondi test is grey (frequency-independent), so the integrated luminosity is independent of the number of bins. By varying the bin number, we are strictly performing a timing comparison when the multi-frequency capabilities of the code are exercised.

An important consideration for the IMC treatment is that the number of packets employed in any given problem and at any given time can be dialed up or down to achieve desired levels of noise. Additionally, the number of active packets changes over the course of the calculation. The average number of active packets that is reported in table~\ref{tab:performance} is calculated based on the number of active packets at the end of each hydro time step over the course of the calculation. This number generally depends on the problem, and on the time at which we terminate the calculation. For the 1D tests, which used 128 zones, the average number of packets per zone works out to about 41000, 130000, and 77000 for cases 3, 4, and 5, respectively, in direct proportion to the performance metric in table~\ref{tab:performance}. For the 3D Bondi test, which used $72^3$ zones, the average number of packets per zone is approximately 380.

We find IMC to be about 1000 times more expensive than the 1D single-group M1 tests, given our packet counts (which we note can be reduced to make IMC more competitive while still maintaining reasonable noise levels). In 3D, the relative performance of the IMC improves significantly due to its intrinsic 3D nature. Depending on the number of M1 frequency groups employed, IMC can actually out-perform M1 on this problem, while maintaining an average of approximately 380 packets per zone. A crucial point here is that, for a fixed level of frequency-integrated sampling noise, the expense of the IMC calculation is effectively insensitive to the number of frequency groups used. If, instead, one wishes to maintain a fixed level of sampling noise in each frequency bin, then the cost of IMC increases linearly with the number of frequency bins, absent further optimizations.

\begin{figure*}[htb!]
  \begin{center}
    \includegraphics[width=0.5\textwidth]{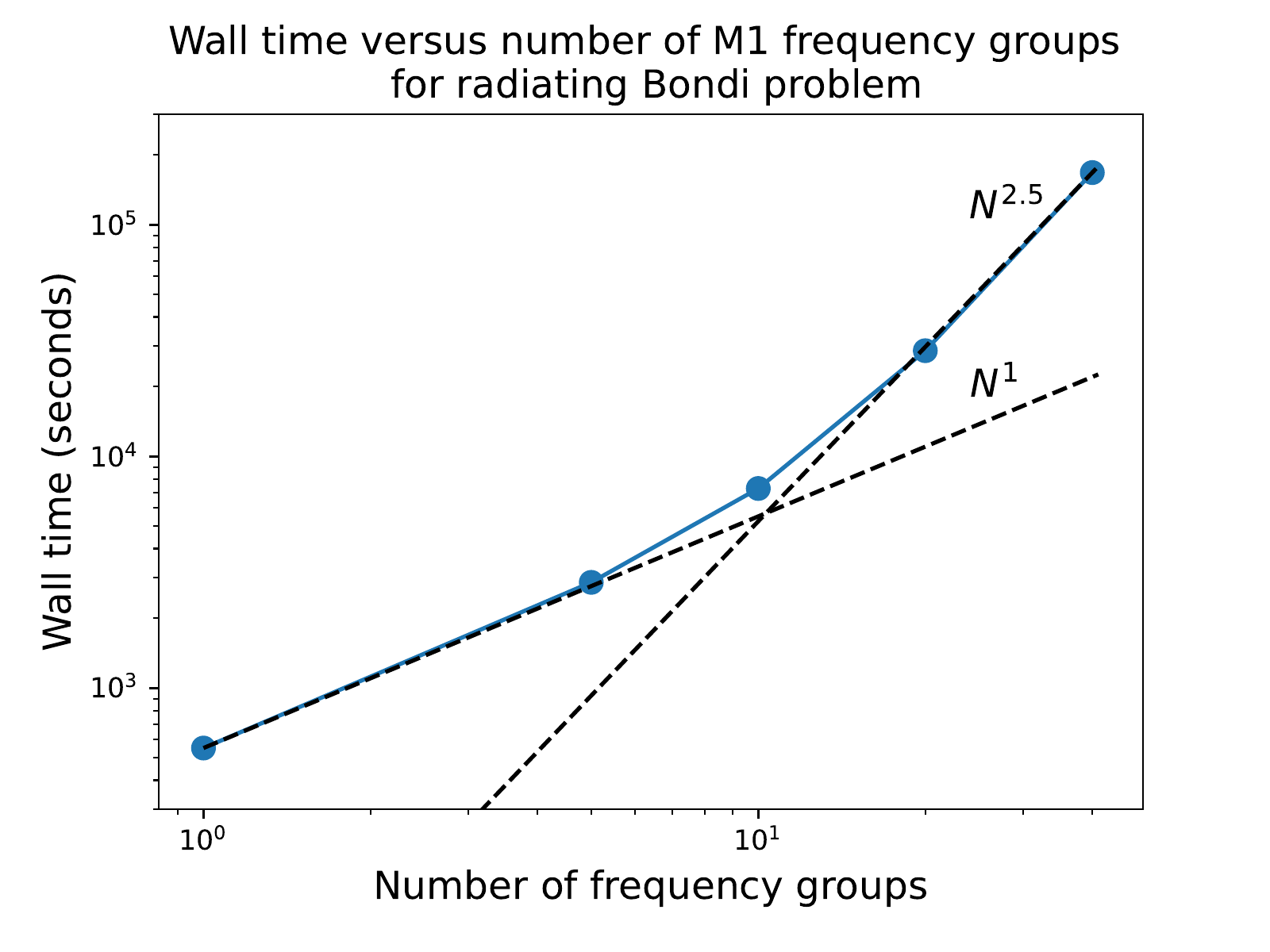}
    \end{center}
\caption{The time required to complete the radiating Bondi problem with M1 transport as a function of frequency bins. All calculations used a spatial resolution of $72^3$ zones, and were parallelized across 64 processors with a $ 4 \times 4 \times 4$ domain decomposition pattern. The 1, 5, 10, and 20 group calculations were run to $t = 5000$ mass units. The 40 group calculation was only run to $t = 1000$, and the time to run to $t=5000$ was estimated based on its average performance to that point.}
\label{fig:M1GroupScaling}
\end{figure*}

By contrast, the expense scaling of M1 with the number of frequency bins can be super-linear. Figure~\ref{fig:M1GroupScaling} shows how the time to complete the radiating Bondi calculation depends on the number of frequency groups. At low group-count the scaling is roughly linear. However, as the number of groups increases, the scaling becomes dominated by the expense of solving a nonlinear $5+4N$ dimensional  matrix system (combined with the primitive inversion elements discussed in section~\ref{subsec:primitive}, and in more detail in \cite{Anninos2020}), where $N$ is the number of frequency groups and the factor of four accounts for energy and flux vector coupling. At 40 groups we observe $N^{2.5}$ scaling.  With even more groups, we expect this scaling to approach $N^3$, the theoretical operational limit of the underlying LU decomposition algorithm used to solve these systems of equations.

\begin{figure*}[htb!]
  \begin{center}
    \includegraphics[width=0.5\textwidth]{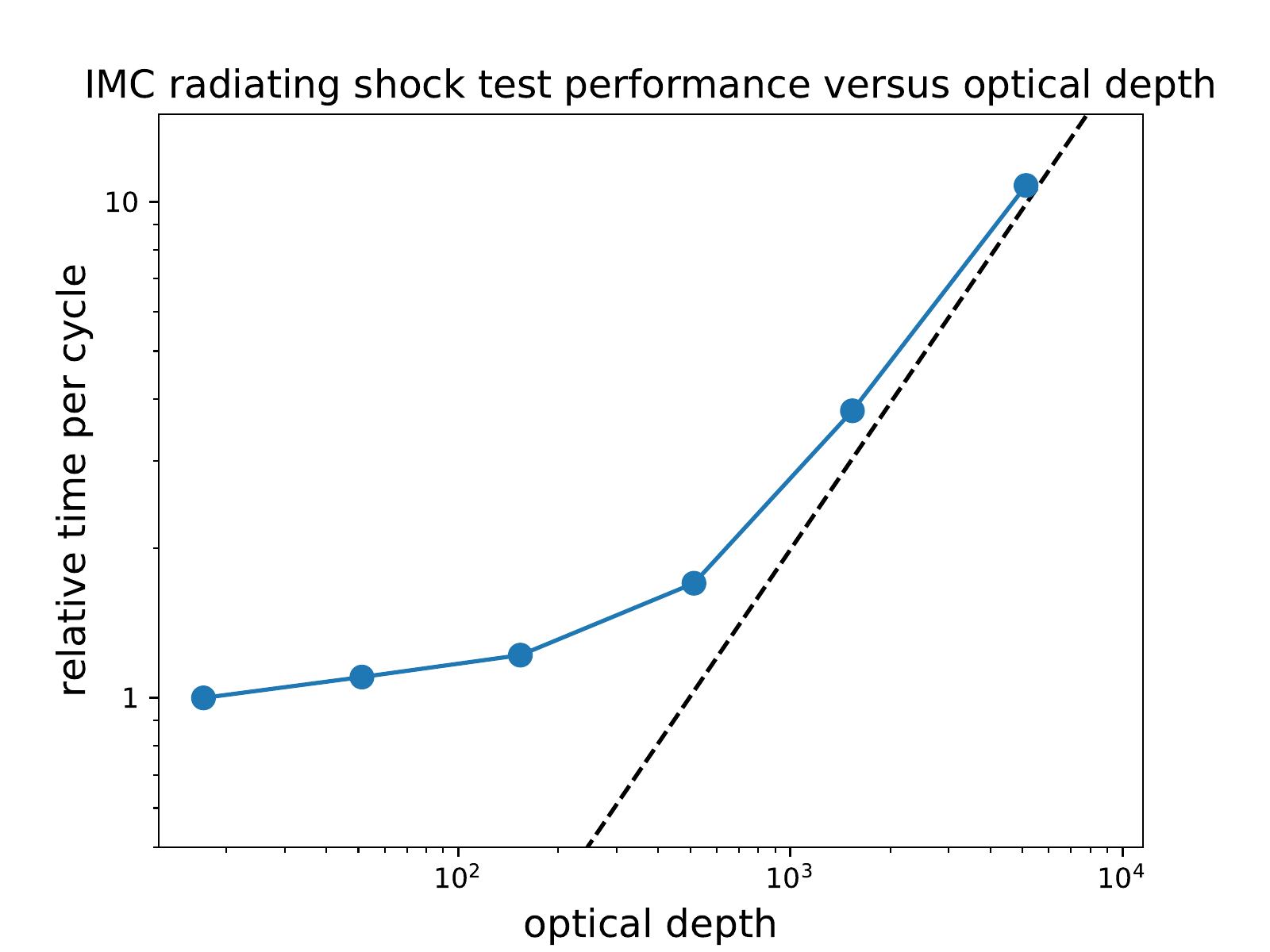}
    \end{center}
\caption{The relative computational expense for the relativistic radiating shock tests as a function of optical depth. These runs are generalizations of cases 4 and 5 from section~\ref{sec:RelativisticRadiatingShocks}, where opacity is the only parameter that is changed between the runs. The optical depth reported on the horizontal axis corresponds to the downstream half of the domain. When the optical depth is high enough such that the average number of packet interactions per time step significantly exceeds unity, the expense scales linearly with optical depth, as indicated by the dashed line.}
\label{fig:IMCKappaScaling}
\end{figure*}

Figure~\ref{fig:IMCKappaScaling} shows how the computational expense of IMC scales with optical depth. The tests used to make this figure are generalizations of the radiation shock tube cases 4 and 5 as described in section~\ref{sec:RelativisticRadiatingShocks}, where only the opacity is changed between calculations. For all the data points shown, the timing measurement is performed once the system has reached steady-state, and with approximately the same number of active packets in all cases. At low enough optical depth, the cost is nearly independent of optical depth because the average number of packet-matter interactions per time step is less than unity, and the packet interactions are the most computationally expensive part of the algorithm. When the optical depth is increased to where packets interact many times per time step, the computational expense approaches a linear scaling as indicated by the dashed line. The fact that computational expense increases for problems with higher optical depths is a well-known feature of a straightforward IMC solution, although this effect may be mitigated with specialized techniques as we discuss in the conclusion.

\section{Conclusions}
\label{sec:Conclusions}

We have expanded the modeling capabilities of our radiation-magnetohydrodynamics code {\em Cosmos++} 
to include implicit Monte Carlo radiation transport that is valid to all orders of ($v/c$), and equally applicable
to Newtonian, special relativistic, and general relativistic problems. This latest capability represents a major
improvement in fidelity over our previous multi-group transport implementation based on an Eulerian 2-moment (M1) closure method.
The two approaches (M1 versus IMC) offer uniquely distinct advantages and disadvantages, so having the ability
to choose one over the other is an important feature of {\em Cosmos++}. Monte Carlo methods, for example,
can be prohibitively expensive at low noise levels, else suffer the consequences of random noise fluctuations. But they generally treat radiation-matter interactions
(e.g., Comptonization) with greater accuracy and spectral fidelity. Zero or first moment multi-group methods, on the other hand,
can be significantly faster without the worry of noise fluctuations when a moderate number of frequency bins will do. But they suffer
in the treatment and accuracy of radiation-matter coupling and anisotropic stress energy closures.
For problems where both spectral fidelity and computational cost are valued, one can imagine performing
a self-consistent RMHD calculation with M1 closure to get reasonable energy balances, then running the Monte Carlo
as a post-processor to resolve the spectrum, an option made available in our implementation.

It is well-known that Monte Carlo techniques are not well-suited for problems in the strong diffusion regime, becoming inefficient
when the algorithms struggle to resolve an increasing number of scattering events at very short spatio-temporal scales.
For explosive systems, this can be dealt with by using the diffusion approximation until the system relaxes to reasonable optical depths. Supernova calculations, for example, activate transport a day or two following the explosion when the optical depth drops below $\sim 10^4$ \citep{Lucy2005, Kasen2006}. In other cases, such as black hole accretion disks, the optical
depth is a fixture in the equilibrium structure of the disk, in which case the diffusion limit has to be dealt with in 
a more rigorous manner. We propose to do that in future work by incorporating random walk algorithms along the lines of \citep{Fleck1984} and its relativistic extension \citep{Richers2017}, Implicit Monte Carlo Diffusion \citep{Gentile2001}, or Discrete Diffusion Monte Carlo \citep{Densmore2007}.

\section*{Acknowledgments}
We thank Nick Gentile for helpful discussions concerning his adaptation of the Fleck factor and photon energy and direction sampling.

This work was performed under the auspices of the U.S. Department of Energy by Lawrence Livermore National Laboratory under Contract DE-AC52-07NA27344.

\end{document}